\documentclass[12pt,preprint]{emulateapj}
\usepackage{apjfonts}

\pdfoutput=1

\usepackage{epsfig}
\usepackage{amsmath}
\usepackage{natbib}

\newcommand\xv{\mbox{\boldmath $x$}}
\newcommand\nablav{\mbox{\boldmath $\nabla$}}
\newcommand\uv{\mbox{\boldmath $u$}}

\newcommand\bvec{\mbox{\boldmath $b$}}

\newcommand\kv{\mbox{\boldmath $k$}}
\newcommand\Hunits{\mbox{km s$^{-1}$ Mpc$^{-1}$}}
\newcommand\ML{\Upsilon_{\rm F606W}}
\newcommand\Rein{R_{ein}}

\newcommand\hacs{{$H_0=85^{+14}_{-13}$ \Hunits} (68\% CL)}
\newcommand\hml{{$H_0=79.3^{+6.7}_{-8.5}$ \Hunits} (68\% CL)}

\begin{document}

\title{Improved Constraints on the Gravitational Lens Q0957+561. \\ II. Strong Lensing}
\author{
  R. Fadely\altaffilmark{1},
  C. R. Keeton\altaffilmark{1},
  R. Nakajima\altaffilmark{2,3},
  and G. M. Bernstein\altaffilmark{2}
}

\altaffiltext{1}{
Department of Physics and Astronomy, Rutgers, The State University of New Jersey,
Piscataway, NJ 08854;
fadely, keeton@physics.rutgers.edu
}

\altaffiltext{2}{
Department of Physics and Astronomy, University of Pennsylvania,
Philadelphia, PA 19104;
garyb@physics.upenn.edu
}

\altaffiltext{3}{
Space Sciences Laboratory, University of California, Berkeley,
Berkeley, CA 94720;
rnakajima@berkeley.edu
}

\begin{abstract}
We present a detailed strong lensing analysis of an HST/ACS legacy dataset for the first gravitational lens, Q0957+561.  With deep imaging we identify 24 new strongly lensed features, which we use to constrain mass models.  We model the stellar component of the lens galaxy using the observed luminosity distribution, and the dark matter halo using several different density profiles.  We draw on the weak lensing analysis by \citet{Nakajima} to constrain the mass sheet and environmental terms in the lens potential.  Adopting the well-measured time delay, we find \hacs\ using lensing constraints alone.  The principal uncertainties in $H_0$ are tied to the stellar mass-to-light ratio (a variant of the radial profile degeneracy in lens models).  Adding constraints from stellar population synthesis models, we obtain  \hml.  We infer that the lens galaxy has a rising rotation curve and a dark matter distribution with an inner core.  Intriguingly, we find the quasar flux ratios predicted by our models to be inconsistent with existing radio measurements, suggesting the presence of substructure in the lens.
\end{abstract}

\section{Introduction}

Since its inception, strong gravitational lensing has been used to probe galaxy mass distributions in a way that complements photometric and dynamical studies. In systems with numerous lensed features, especially partial or full Einstein rings, lensing can provide a fairly detailed description of the lens galaxy mass distribution \citep[e.g.,][]{RingCycle,KKM,Kochanek95,Lehar96,Lehar97,Trotter,Koopmans,Gavazzi,Suyu1,Suyu2}. In systems with only a few lensed images, however, there may be significant systematic uncertainties in the lens mass distribution. For example, four-image lenses typically constrain the angular structure of the lens potential reasonably well \citep[e.g.,][]{Keeton97b}, but often cannot determine the radial profile because the images lie at similar distances from the center of the galaxy \citep[e.g.,][]{Keeton97a,Kochanek02}.  Two-image lenses are better able to probe the radial profile, but provide only poor constraints on the angular structure of the lens potential \citep{Rusin03}.  In addition, further complications may arise if the lens galaxy lies in a group or cluster of galaxies that significantly affect the lens potential \citep[see][]{Keeton04}.

These issues explain the challenges associated with modeling the first gravitational lens discovered, Q0957+561 \citep{Walsh}. The two lensed images of the background quasar provide limited constraints on a lens potential that is complicated by the presence of a cluster surrounding the main lens galaxy  \citep[e.g.,][]{Kochanek91}.  To move beyond simple two-image reconstructions, \citet{GorensteinVLBI} and \citet{Garrett} used VLBI observations to resolve the radio structure of the quasars; but \citet{Grogin} and \citet{Barkana} found that the new radio constraints still could not strongly constrain the lens potential. Seeking yet more constraints, \citet{Bernstein97} and \citet{Keeton00} used the Hubble Space Telescope to discover lensed images of the quasar host galaxy, but even they found that a range of models could reproduce all the strong lensing data \citep[also see][]{Bernstein99}. There were parallel efforts to obtain other kinds of data to provide independent constraints on the lens galaxy and cluster. \citet{Tonry} measured the central velocity dispersion of the lens galaxy, which \citet{Romanowsky} used in stellar dynamical models. \citet{Fischer} studied weak lensing by the cluster. \citet{Chartas} used the Chandra X-ray Observatory to detect the hot intracluster gas and estimate the cluster mass. Despite the growing amount of data for this system, a definitive description of the mass distribution has remained elusive.  

\begin{figure}[t]
\centering
\includegraphics[clip=true, trim=1.6cm -0.50cm 0cm 0cm,width=11cm]{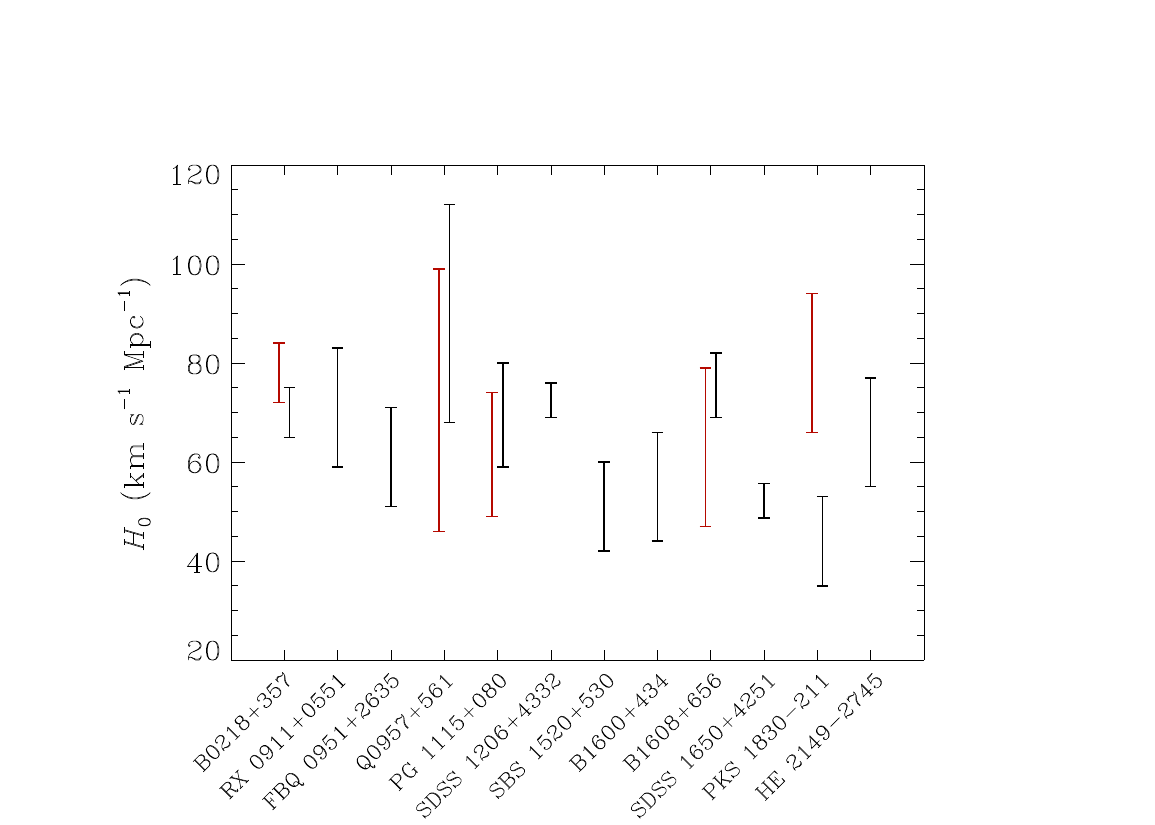}
\caption{Current $H_0$ measurements in individual lens studies. Black errorbars indicate the most recent measurement for a given lens.  When more than one such effort has been made, we present the second most recent effort in red.  The measurements span a range of median values and uncertainties; a simple $\chi^2$ test indicates the scatter is not purely statistical at $>99$\% confidence.
\textit{References---} B0218+357: \citet{Wucknitz, York}, RX J0911+0551: \citet{Hjorth}, FBQ0951+2635: \citet{Jakobsson}, Q0957+561: \citet{Bernstein99, Keeton00}, PG 1115+080: \citet{Treu02, Chartas04}, SDSS 1206+4332: \citet{Paraficz}, SBS 1520+530: \citet{Burud02b}, B1600+434: \citet{Koopmans00, Burud00}, B1608+656: \citet{Fassnacht02, Koopmans03}, SDSS 1650+4251: \citet{Vuissoz}, PKS 1830-211: \citet{Lidman,Winn},  HE 2149-2745: \citet{Burud02a}. }
\label{fig:hlenses}
\end{figure}

Uncertainties in the mass distribution propagate into attempts to use the lens time delay to determine the Hubble constant, $H_0$. In principle, lens time delays offer a simple measurement of the Hubble constant at cosmological distances that bypasses the distance ladder. In practice, however, $H_0$ is degenerate with certain aspects of the mass model \citep{Falco85,Gorenstein,Williams00,Kochanek02,Saha06}.  Due to these degeneracies, measurements of $H_0$ in individual lens systems have yielded quite varied results.  Figure \ref{fig:hlenses} shows the lens systems for which individual modeling of the mass distribution has been done to determine $H_0$.  Examining the figure, it is unclear what value of $H_0$ such studies prefer.  Roughly half of the previous studies are consistent with $H_0<60\ \Hunits$, with most preferring $H_0=65$--$80\ \Hunits$.  Strikingly, only two studies, of Q0957+561 \citep{Bernstein99, Keeton00} and PKS 1830-211 \citep{Lidman}, found $H_0$ values $>85\ \Hunits$.

In non-lensing studies, measurements of $H_0$ have reached better agreement.  Using Cepheid variable stars as distance indicators, \citet{Freedman} obtained $H_0=72\pm2$ (stat.) $\pm7$ (syst.) \Hunits.  \citet{Riess} recently analyzed a higher quality and more homogeneous sample of Cepheids and supernovae and found $H_0=74.2\pm3.6\ \Hunits$.  Results from measurements of the Cosmic Microwave Background (CMB) have given similar results.  Most recently, \citet{Dunkley} analyzed the 5th year data release of the Wilkinson Microwave Anisotropy Probe (WMAP) to find $H_0=71.9^{+2.6}_{-2.7}$ \Hunits, assuming a universe with a flat geometry and a cosmological constant.  With joint analyses of the various constraints, measurements of $H_0$ can now approach the percent level.  For instance, \citet{Komatsu} combined $H_0$ measurements from the CMB, SNIa, and Baryon Acoustic Oscillations (BAO) to obtain $H_0=70.5\pm1.3\ \Hunits$ or a 2.3\% determination of $H_0$.

Given the apparent consensus in non-lensing studies of $H_0$, the scatter in results from lensing is puzzling.  Are conventional lens models failing to account for variations in key components of lens models, such as the logarithmic density slope, angular structure, and/or substructure?  Are the (often large) uncertainties in lensing measurements of $H_0$ fully understood?  How can the uncertainties be reduced?

One way to confront these concerns is to model a statistical ensemble of lenses when measuring $H_0$.  \citet{Saha06} used non-parametric lens modeling to generate free-form mass maps of 10 lenses and obtain $H_0=72.5^{+7.8}_{-11.3}\ \Hunits$.  Adding one lens to this sample, \citet{Coles} used similar techniques and found $H_0=71^{+6}_{-8}\ \Hunits$.  \citet{Oguri} took a parametric approach but attempted to incorporate a reasonable amount of complexity and scatter in the models; his analysis of 16 time delay lenses yielded $H_0=68\pm6$ (stat.) $\pm8$ (syst.) \Hunits.  While these results are enticing, the systematic uncertainties may still be underestimated and cannot be beaten down with sample size.  In particular, lensing constraints on $H_0$ are known to be affected by the ``mass-sheet degeneracy'': a uniform sheet of mass can be added to a lensing potential in a way that leaves the positions and brightnesses of the images unchanged but rescales the inferred Hubble constant \citep{Falco85,Gorenstein}.  Groups and clusters surrounding lens galaxies or along the line of sight can act as mass sheets that bias lensing measurements of $H_0$ \citep{Keeton04}.  While there is considerable effort to characterize the local and line-of-sight environments of strong lenses \citep[e.g.,][]{Fassnacht-0712,Fassnacht-1608,Fassnacht-Xray,Iva,Kurtis,Auger}, and \citet{Oguri} attempted to account for the mass sheet in several lenses for which it is expected to be significant, it is still not clear whether the systematic uncertainties are fully understood.

In order to identify and control systematics, it seems that we still need to model individual lens systems in detail.  One advantage of this approach is the ability to use additional sources of information, such as weak lensing \citep[e.g.,][]{Fischer,Bernstein99} or stellar dynamics \citep[e.g.,][]{Grogin,Tonry,Treu02,Barnabe07}, to reduce the uncertainties inherent in strong lensing.  With the four-image lens B1608+656 \citep{Myers}, for example, \citet{Suyu1,Suyu2} demonstrate that lensing and velocity dispersion information can be combined to get a $\sim7\%$ determination of $H_0$, in which the modeling uncertainties are comparable to the uncertainties in the measured time delays themselves \citep{Fassnacht02}.  With Q0957, we have a system whose complicated lens potential not only presents certain challenges for measuring $H_0$, but also bestows certain opportunities.  In particular, the presence  of the cluster around the lens galaxies offers a rare chance to do a weak lensing analysis of an \emph{individual} strong lens system (as opposed to a stacked ensemble; e.g., \citealt{Gavazzi-WL,lagattuta}), which is key to breaking the mass sheet degeneracy and controlling that vital systematic uncertainty.  Furthermore, there is now a rich variety of data available for Q0957, which enables a broad-based analysis.  We therefore study Q0957 to investigate statistical and systematic uncertainties in lens models and $H_0$.  The strong lensing analysis presented here complements and draws upon the weak lensing analysis presented by \citet{Nakajima}; both are based on the same new HST/ACS data.  In this system extensive observational effort \citep[e.g.,][]{Schild,Kundic,Oscoz} has yielded a very precise time delay with an uncertainty of just $\sim0.02\%$ \citep{Colley}. While we do not expect to achieve that level of precision in lens models, our goal is to see just how well we can do.  We combine our new data with a number of independent constraints from previous observations as well as stellar population synthesis models, in order to obtain the most precise measurement of $H_0$ to date for Q0957.

In this paper we assume a flat universe with matter density $\Omega_M=0.274$ and cosmological constant $\Omega_{\Lambda}=0.726$, which is consistent with the 5-year WMAP+SN+BAO constraints from \citet{Komatsu}.

\begin{figure*}[!ht]
\centering
\includegraphics[clip=true, trim=0.9cm 5cm 2.825cm 5.05cm, height=3.25in, angle=270]{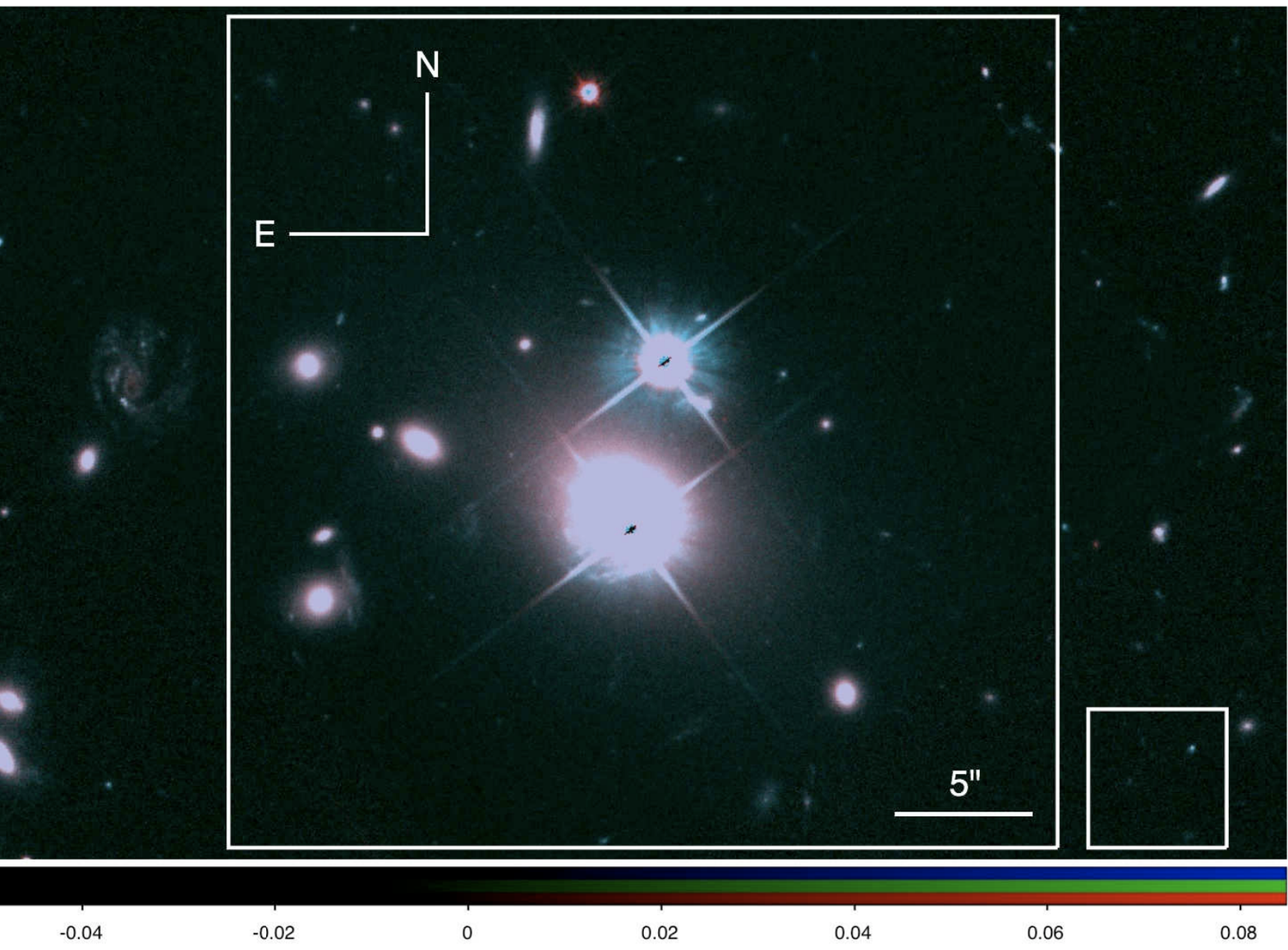}
\includegraphics[clip=true, trim=1.375cm 6.6cm 1.6cm 7.60cm, height=2.395in, angle=270]{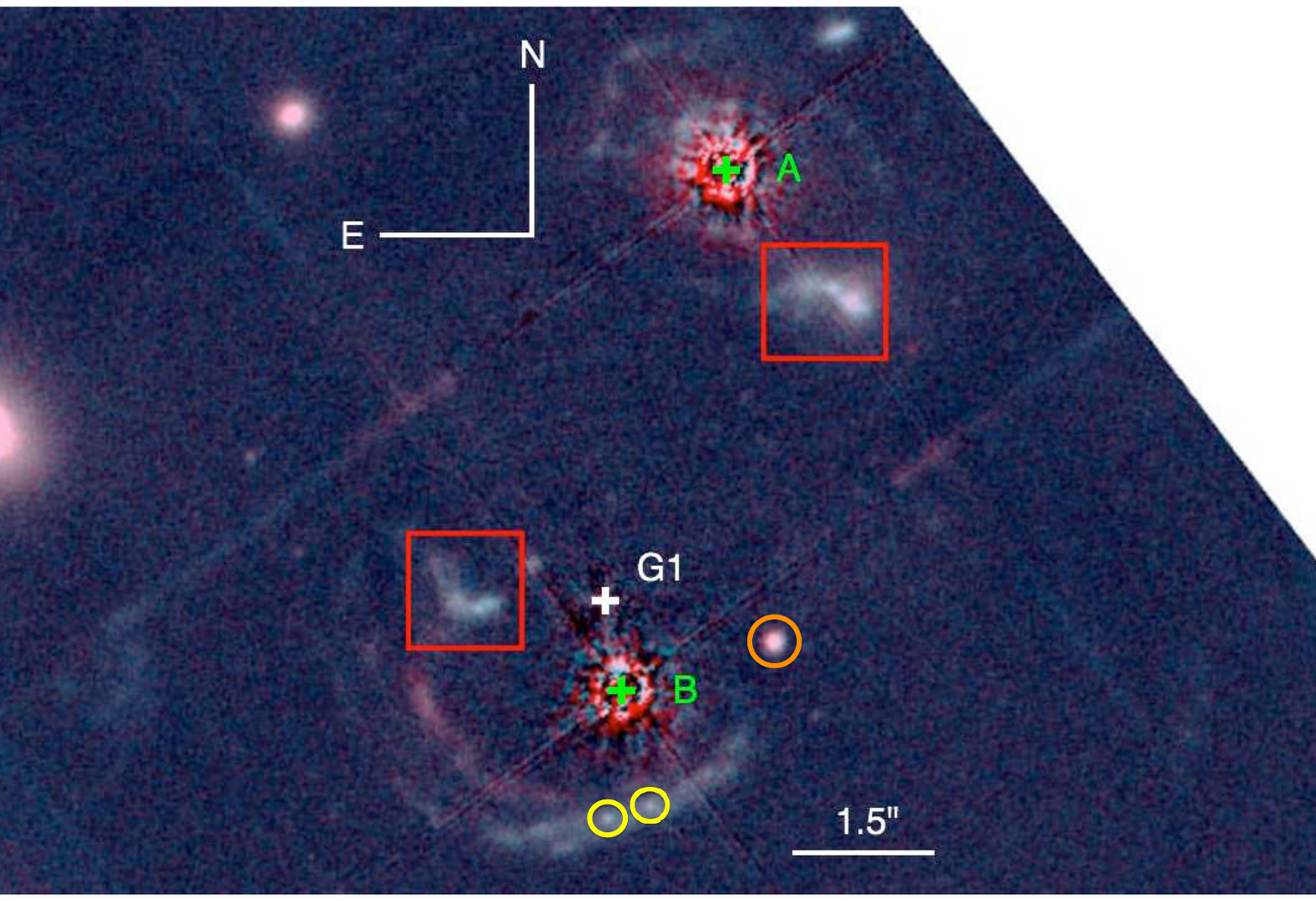}
\caption{
\textit{(a, left)} A false color image of the central $30''$ of our combined F606W and F814W images of Q0957+561.
\textit{(b, right)} Close-up of the strong lensing region, after the main lensing galaxy and quasar images have been subtracted.  The white cross indicates the center of the lens galaxy G1, while the green crosses indicate the quasar positions A and B. The red boxes and yellow circles indicate the ``blobs'' and ``knots'' identified by \citet{Bernstein97}.  Newly resolved faint features are seen south and east around quasar B and southwest of quasar A.  The orange circle indicates an unknown object, not associated with any lensed features.  Since the light profile of the object is consistent with the PSF, we surmise it is a faint halo star in the foreground of the lens.}
\label{fig:picture}
\end{figure*}

\section{Observations and Data Analysis}
\subsection{Observations}
\label{sec:obs}

We conducted deep observations of Q0957+561 on 2005 October 10--11 with HST's Advanced Camera for Surveys as part of program GO-10569.  Using four pointings of 7.7 ks in the F606W filter and 3.8 ks in the F814W filter, we created a $6'\times6'$ mosaic of the field.  We arranged the pointings to overlap in the central 30$''$ region, forming a 30 ks image in the F606W filter (15 ks in F814W) for our strong lensing analysis with a final pixel scale of 0.03$''$.  The large number of exposures in this central region allows us to use a simple image-combining algorithm that avoids the undesirable PSF broadening and noise correlation of the common \textit{Drizzle} algorithm \citep{FruchterHook}:
\begin{enumerate}
\item An astrometric solution is derived for each exposure by compounding the ACS/WFC coordinate map of \citet{Anderson} with an additional affine transformation to account for pointing errors, stellar aberration, and slight plate-scale variations due to the HST ``breathing mode.''  The coefficients of the affine transformation are derived by registering objects detected in individual exposures.
\item A grid of $0.03''$ pixels is created for the combined image.  Each pixel in each exposure is mapped to a single destination pixel.  Input pixels flagged as invalid due to detector defects, etc., are discarded.
\item For each destination pixel, we average all of the input pixels, using a sigma-clipping algorithm to eliminate pixels contaminated by cosmic rays.
\end{enumerate}
The procedure is identical to the use of \textit{Drizzle} with a ``drop zone'' of zero size.  Since each input pixel contributes to only one output pixel, the output pixels have uncorrelated noise.  The combining algorithm broadens the PSF only by an effective convolution with the output pixel square.  The final pixel size is chosen such that it is small enough not to degrade the resolution, but coarse enough that there are enough input pixels for averaging and outlier rejection.  We present a false color image of our combined F606W and F814W images in Figure \ref{fig:picture}a.

To look for new strong lensing constraints, we subtract the bright quasar images A and B using the PSF derived from observations of the star HD237859.  Since the PSF varies with focal plane position as well as time, we observed the star as close as possible in time and chip position to each quasar image in each of the four pointings.

\subsection{Lens Galaxy Properties}
\label{sec:galaxy}

\begin{figure*}[!ht]
\centering
\includegraphics[width=8cm]{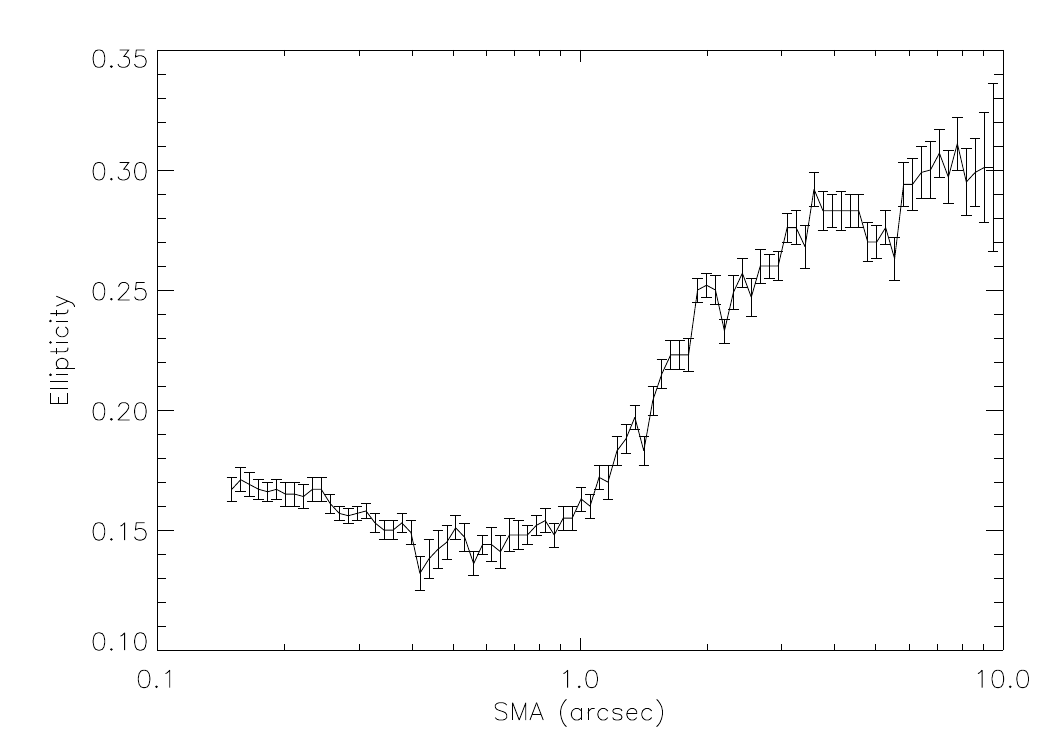}
\includegraphics[width=8cm]{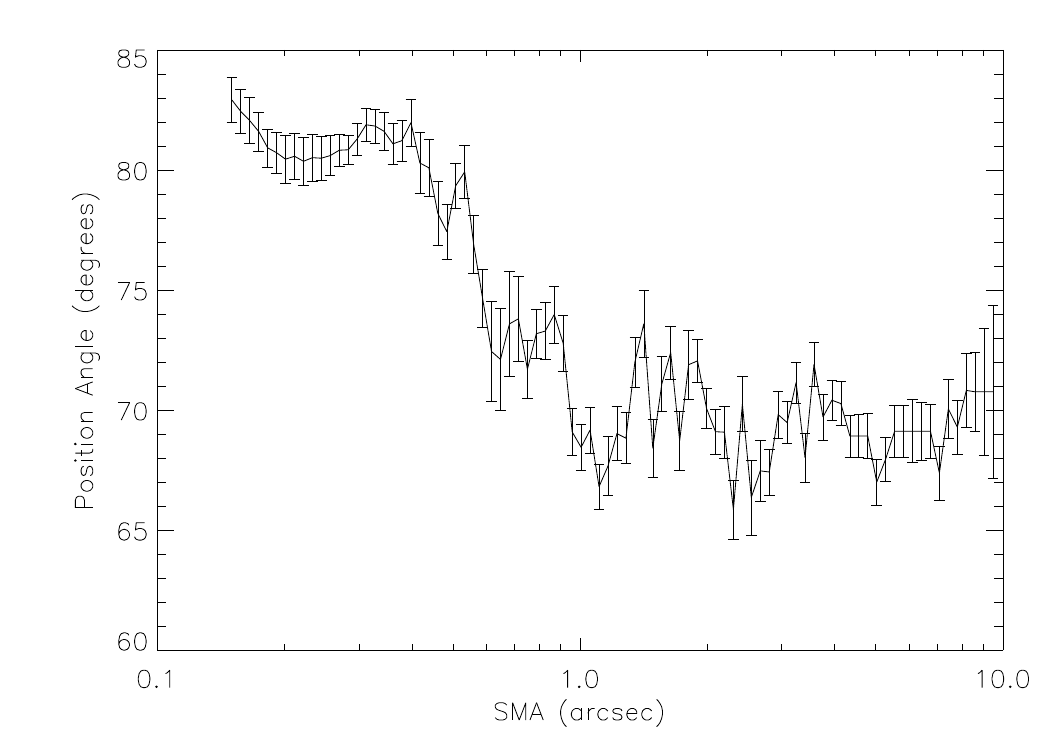}
\caption{Ellipticity and position angle of galaxy isophotes, plotted as a function of semi-major axis.  Note the increase in ellipticity beginning at $\sim 1''$ and the decrease in position angle beginning at $\sim 0.3''$.}
\label{fig:e-and-pa}
\end{figure*}

We model the main lensing galaxy using the IRAF ELLIPSE task, masking out regions where quasar subtraction takes place as well as any bright regions not associated with the galaxy (e.g., other lensed features).  Our resulting IRAF model provides a measurement of the galaxy's isophotes and total flux (see Table \ref{tab:galphot}).  As shown previously \citep{Bernstein97}, the isophotes of the lens galaxy exhibit an ellipticity gradient and a position angle twist (Figure \ref{fig:e-and-pa}).  These isophotal features may complicate the lensing potential, so we incorporate them directly into our lens models (\S \ref{sec:models}).  We also use the photometry of the lens galaxy to constrain stellar population synthesis models and estimate the stellar mass-to-light ratio (\S \ref{sec:SPS}).

\subsection{Faint Strong Lensing Features}
\label{sec:features}

We subtract the model galaxy from the quasar-subtracted image to produce the final image of the strong lensing region, which is shown in Figure \ref{fig:picture}b.  This image reveals several new, previously unresolved or undetected strongly lensed features.  Since the morphology is similar to the host galaxy arc from NICMOS \citep{Keeton00}, we conjecture that the optical features are most likely images of star forming regions of the quasar host galaxy at $z=1.41$.

The lensed ``blobs'' and ``knots'' indicated in Figure \ref{fig:picture}b were previously identified by \citet{Bernstein97}, and were used by \citet{Bernstein99} and \citet{Keeton00} as constraints on lens models.  To derive new constraints from our new strongly lensed features, we use the models of \citet{Keeton00} as a starting point.  Using the \textit{lensmodel} software \citep{Keeton01}, we check to see how these older models would map new features in the image plane.  Specifically, we take an observed image position, map it to the source plane, and then find all corresponding images using the old lens models.  We then look for the predicted images in our new HST data.  Unfortunately, we find the \citeauthor{Keeton00} models cannot sensibly reproduce the lensing we see in the HST data.  These models fail most notably for the new features south and east of quasar B, mapping bright peaks in the data to blank regions of the sky.

\begin{figure*}[ht]
\centering
\includegraphics[clip=true, trim=4cm 8cm 7cm 7cm, height=9cm]{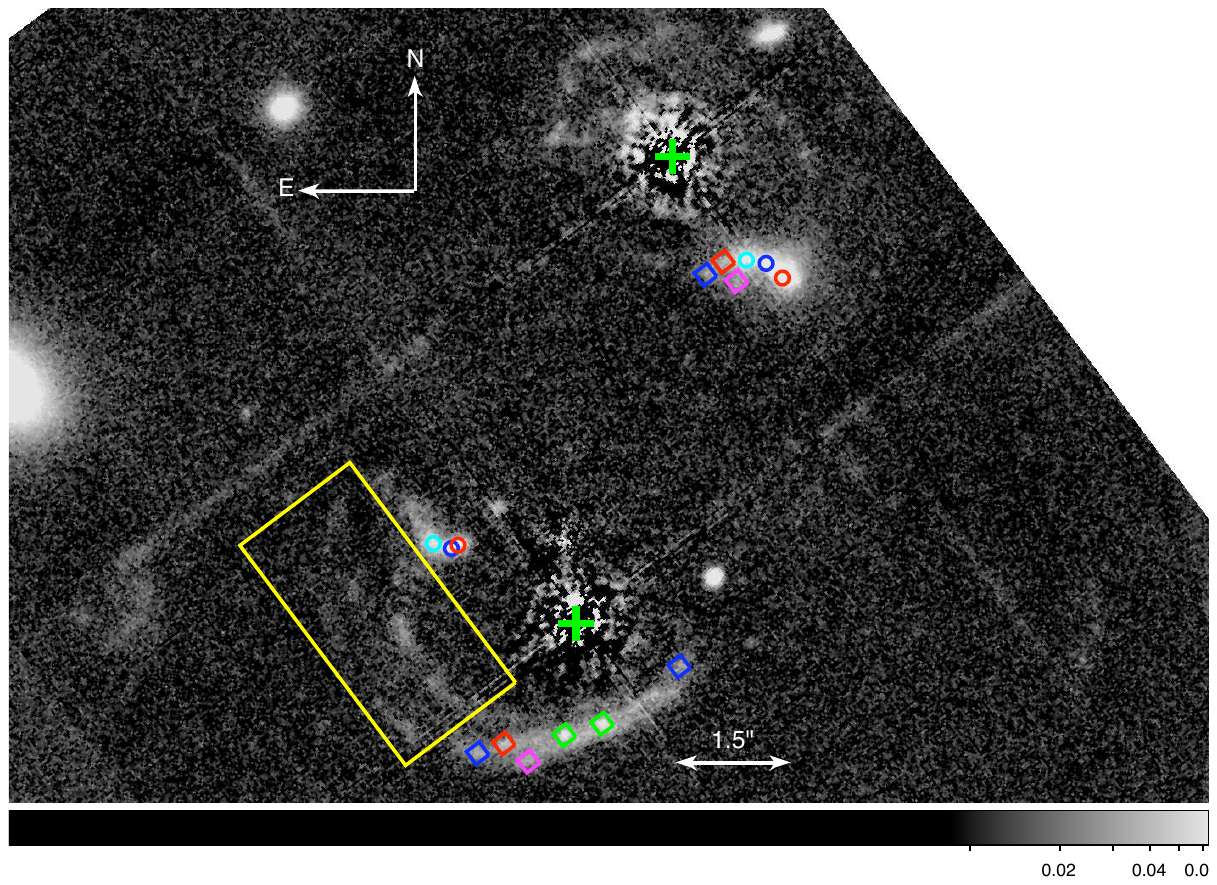}
\includegraphics[clip=true, trim=4cm 8cm 7cm 7cm, height=9cm]{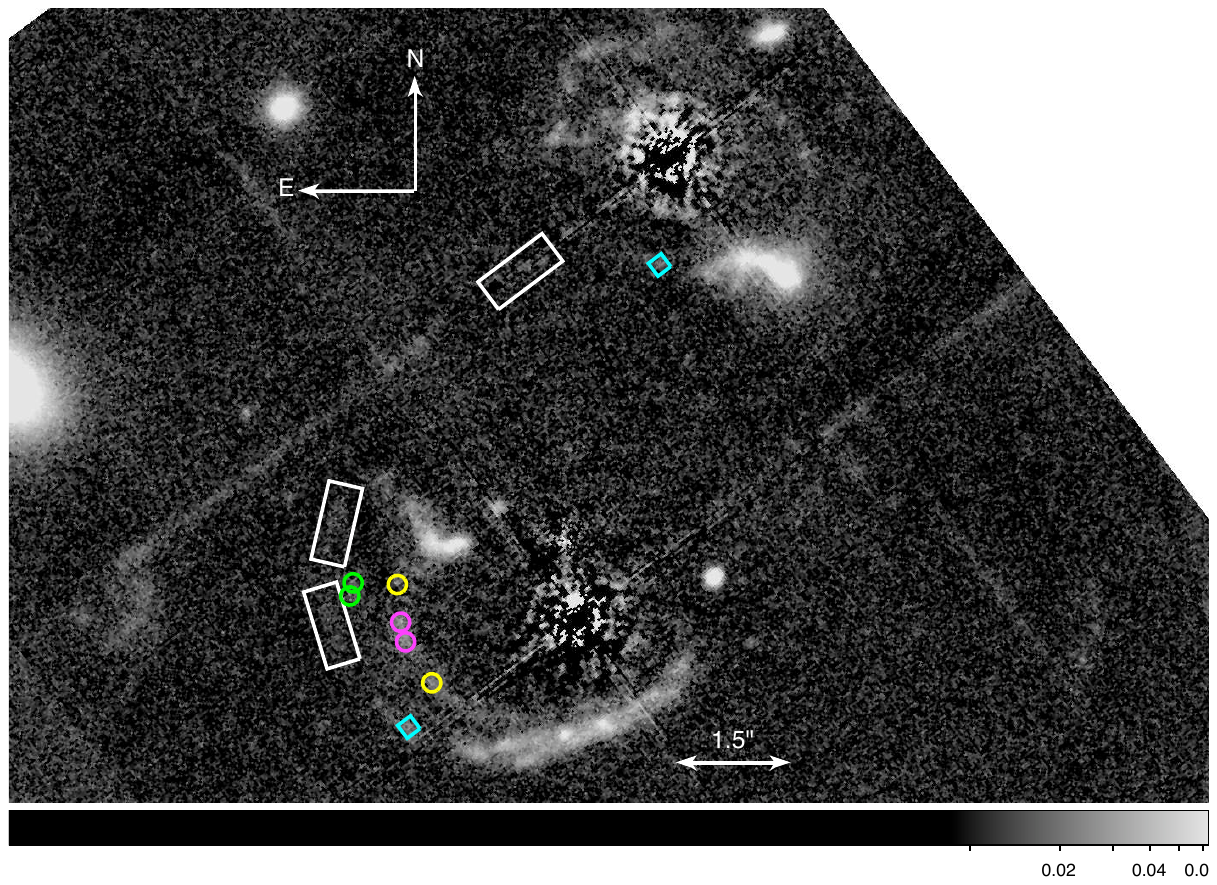}
\caption{
Our proposed mapping of the new lensed features (cf.\ Table \ref{tab:data}).  Corresponding shapes and colors indicate proposed images of a given source.   The grid-like pattern that appears is due to the imperfect subtraction of the diffraction pattern of the quasar images.
\textit{(a, left)} We use the images indicated to constrain new lens models, and then use those models to test features in the yellow box.
\textit{(b, right)} We take one of the points in each shape/color pair, map it back to the source, and then find the corresponding images. The predicted positions match very well with features in the image, which gives us confidence both that the ``primary'' constraints are valid and that the additional lensed features are real.  (Points in white rectangles indicate features that are too close to the noise level to provide confident peaks.)  All of the colored points are used as constraints in our subsequent modeling.}
\label{fig:new}
\end{figure*}

In order to make sense of the new features, we examine the morphology of the images in the data.  Specifically, in the area around the bright ``knots'' shown in Figure \ref{fig:picture}b we notice a distinct fork-like feature extending from either side of the knots.  Using peaks in this structure, we postulate a set of new constraints as shown in the left panel of Figure \ref{fig:new}, which we call our ``primary'' set of new constraints.  We use these primary features to constrain a singular isothermal ellipsoid lens model, and see how the resulting model maps other faint features found in the HST data.  Specifically, we consider faint features to the east of quasar B, indicated by the large box in Figure \ref{fig:new}a.  To our surprise, the model derived from our primary features accurately describes the remaining faint features, mapping peaks of the images in reasonable ways (Figure \ref{fig:new}b) and giving us confidence in the identification of new lensed features.  We therefore add these features to our list of constraints, resulting in a final data set which is given in Table \ref{tab:data}.  We note that all presumed multiple images of each source are observed to have similar colors (within the photometric noise).

To obtain the position and uncertainty of each peak listed in Table \ref{tab:data}, we examine the brightest pixel(s) within the peak.  If the brightest pixel is more than $10\sigma_{noise}$ above any other pixel in the peak (as for the ``knots'' or source IV in Table \ref{tab:data}), we set the position error to be $\pm1$ pixel or $\pm0.03''$.  If the brightest pixel is 3--$10\sigma_{noise}$ above the surrounding pixels, we conservatively set the error to $\pm1.5$ pixels or $\pm0.05''$.  If there are multiple pixels in the peak that are within $3\sigma_{noise}$ of the brightest, we take the average position of all such pixels and set the error to be the distance from this average to the farthest of the bright pixels, plus our nominal value of $0.05''$.

\section{Lens Modeling Methods} 
\label{sec:modeling}
\subsection{General Theory}
\label{sec:theory}

In this section we present a brief review of the lensing theory that is particularly pertinent to our analysis of Q0957.  For further discussion of strong lensing theory, please see \citet{Schneider} and \citet{Kochanek04}.

As predicted by Einstein's General Relativity in the early 20th century, a mass concentration near the line of sight to a background object may significantly displace and distort a background image.  The angular position $\uv$ of the source and the angular position of $\xv$ of an image are related by the lens equation,
\begin{eqnarray}
\uv=\xv-\nablav \phi\,,
\end{eqnarray}
where $\phi$ is the (scaled) gravitational potential due to mass at the lens redshift.  The lens potential is given by
\begin{eqnarray} \label{eq:Poisson}
\nabla^2\phi (\xv) = 2\kappa (\xv) = 2\Sigma/\Sigma_{crit},
\end{eqnarray}
where the convergence $\kappa$ equals the surface mass density ($\Sigma$) scaled by the critical density for lensing ($\Sigma_{crit}$).  

Since the deflected light travels along different ray paths, there is a difference in the light travel time for different images.  This difference, known as the time delay, can be measured if the source is sufficiently variable.  The time delay between images at positions $\xv_i$ and $\xv_j$ is given by
\begin{eqnarray}
\Delta t_{ij} &=& \frac{1+z_l}{c}\frac{D_{ol}D_{os}}{D_{ls}} \\
&& \times \left\{\frac{1}{2}\left(|\xv_i-\uv|^2-|\xv_j-\uv|^2\right)-\big[\phi(\xv_i)-\phi(\xv_j)\big]\right\} \nonumber ,
\end{eqnarray}
where $z_l$ is the lens redshift and $D_{ol}$, $D_{os}$, and $D_{ls}$ are angular-diameter distances from the observer to the lens, the observer to the source, and the lens to the source respectively.  Combining a measured time delay with a lens model (to infer the source position $\uv$ and the lens potential $\phi$) provides a measurement of the distance combination $D_{ol} D_{os}/D_{ls}$, which is inversely proportional to $H_0$.  (The distance ratio also depends on cosmological parameters $\Omega_M$ and $\Omega_{\Lambda}$, but that is typically a small effect compared with other uncertainties in the problem.)

A well-known problem in lensing constraints on $H_0$ is the ``mass-sheet degeneracy'' \citep{Falco85,Gorenstein}.  For any potential $\phi$ that fits the data, one can construct another potential
\begin{eqnarray}
\phi'=\frac{1}{2}\kappa'|\xv|^2 + (1-\kappa')\phi
\end{eqnarray}
that yields exactly the same image positions and flux ratios.  The addition of the mass sheet $\kappa'$ does, however, rescale the time delays and hence the inferred Hubble constant, such that $H_0'=(1-\kappa')H_0$.  The challenge for Q0957 is that the cluster around the lens contributes a term ($\kappa_c$ in eq.~\ref{eq:cluspot} below) that acts like a mass sheet, which we cannot constrain by fitting the positions and fluxes of the strongly lensed images.  We therefore set $\kappa_c=0$ when doing the strong lens modeling, so we obtain some Hubble constant estimate $H_{0,model}$.  We then use weak lensing data to constrain $\kappa_c$, which yields our corrected Hubble constant estimate
\begin{equation} \label{eq:H0scaling}
  H_0 = (1-\kappa_c) H_{0,model} .
\end{equation}
Additionally, since the mass sheet correction is a rescaling of the potential, we must multiply each term in the potential by the same factor of $1-\kappa_c$ as in eq.~(\ref{eq:H0scaling}).  For our results in \S \ref{sec:results}, we indicate all parameters to which this applies.

\subsection{Mass Models}
\label{sec:models}

\subsubsection{Stellar component}

The lensing potential of Q0957+561 may be complicated by the ellipticity gradient and isophote twist seen in the luminous component of the lens galaxy (see Figure \ref{fig:e-and-pa}).  To allow such features in lens models, \citet{Keeton00} introduced ``double pseudo-Jaffe'' models featuring a superposition of two ellipsoidal mass distributions, centered on the galaxy position, with different ellipticities, orientations, and scale radii. We initially adopted similar models, but quickly judged them to be unsatisfactory.  Constrained by the new lensed features, double pseudo-Jaffe models were adopting odd forms, such as one round and one very flattened component, that seemed unrealistic and unlike what is observed for giant elliptical galaxies.

A better approach is to incorporate the observed ellipticity gradient and isophote twist directly by using our isophotal model of the stellar component.  We combine the model galaxy with an assumed stellar mass-to-light ratio to construct a convergence map.  (We vary the mass-to-light ratio during the modeling, as discussed in \S \ref{sec:SL}.)  To compute the corresponding lensing potential, we solve the Poisson equation using Fourier transforms.  In Fourier space, the Poisson equation (\ref{eq:Poisson}) has the form
\begin{eqnarray}
-\kv^2 F(\kv) = 2K(\kv)
\end{eqnarray}
where $K(\kv)$ is the Fourier transform of the convergence map and $F(\kv)$ is the Fourier transform of the lens potential.  It is straightforward to construct the convergence on a two-dimensional grid, calculate $K(\kv)$ using FFTs, solve for $F(\kv)$, and then do an inverse Fourier transform to get $\phi(\xv)$.  We can then obtain the lensing deflection and magnification by computing derivatives of $\phi(\xv)$ in Fourier space.  This method is discussed in more detail by \citet{vandeven09}.

\subsubsection{Dark matter}

The luminous galaxy is presumably embedded in its own dark matter halo and any halo associated with the surrounding cluster.  The cluster in Q0957 is non-negligible: previous weak lensing studies constrained the mean convergence within $30''$ of the lens to be $\kappa_{30''}=0.26\pm0.16$ and indicated a shear of $\gamma\sim0.2$ \citep{Fischer,Bernstein99}. (In \S \ref{sec:WL} we report our own constraints on cluster mass models.)  Also, X-ray observations of the intracluster gas indicated a convergence from the cluster at the quasar positions of $\kappa_A=0.22^{+0.14}_{-0.07}$ and $\kappa_B=0.21^{+0.12}_{-0.07}$ \citep{Chartas}.

Since the lens galaxy is the brightest and (presumably) most massive galaxy in the cluster, it seems natural to assume as a fiducial model that the galaxy lies at the center of the cluster.  In this case the dark matter halo we insert in our models represents some combination of dark matter associated with the galaxy and dark matter associated with the cluster as a whole.  We consider various profiles to encompass the range of possibilities.  One model we use is the Navarro-Frenk-White profile \citep[NFW,][]{Navarro},
\begin{eqnarray}
\rho=\frac{\rho_s}{(r/r_s)(1+r/r_s)^2}
\end{eqnarray}
whose projected surface density has the form \citep{Bartelmann}
\begin{eqnarray}
\kappa(r)=2\kappa_s\frac{1-\mathcal{F}(x)}{x^2-1}
\end{eqnarray}
where $x=r/r_s$, $\kappa_s=\rho_sr_s/\Sigma_{crit}$, and 
\begin{eqnarray}
\mathcal{F}(x)=\left\{\begin{array}{l l}\frac{1}{\sqrt{x^2-1}}\rm{tan}^{-1}\sqrt{\textit{x}^2-1} & \mbox{($x>1$)}\\ \frac{1}{\sqrt{1-x^2}}\rm{tanh}^{-1}\sqrt{1-\textit{x}^2} & \mbox{($x<1$)}\\ 1 & \mbox{($x=1$)}\\ \end{array} \right. \
\end{eqnarray}
While these equations describe a spherically symmetric model, we can obtain an elliptical model by replacing $r$ with the ellipse coordinate $\xi=(x^2+y^2/q^2)^{1/2}$ where $q$ is the projected axis ratio.  The lensing potential and its derivatives can be computed for elliptical models using numerical integrals \citep{Schramm,Keeton01}. 

We also use three softened power law profiles with projected surface densities of the form
\begin{eqnarray}
\kappa(\xi)=\frac{1}{2}\frac{b_d^{2-\alpha}}{(a^2+\xi^2)^{1-\alpha/2}}
\end{eqnarray}
where $\xi$ is the ellipse coordinate defined above, $a$ is the core radius, $b_d$ is a normalization parameter, and the power law index $\alpha$ is chosen such that $M(R) \propto R^\alpha$ for $R \gg a$. In this class of models we study one with an isothermal profile ($\alpha=1$), one steeper than isothermal ($\alpha=0.5$), and one shallower than isothermal ($\alpha=1.5$).  For $\alpha=1$ the lensing potential and its derivatives can be computed analytically \citep{KK98}, but for $\alpha \neq 1$ they require numerical integrals.

We still need to consider the possibility that the cluster may not be centered on the lens galaxy, which is germane because the observed X-ray emission from the cluster is slightly offset from the lens: the X-ray centroid is $4.3\pm1.3$ arcsec East and $3.2^{+1.2}_{-0.6}$ arcsec North of Image B, or $4.7\pm1.2$ arcsec from the lens galaxy \citep{Chartas}.  We carry out our full modeling analysis (as described in the remainder of the paper) treating the cluster as a distinct dark matter halo centered at the observed X-ray position.  While this approach yields fits that are formally comparable to our fiducial results, we ultimately reject such models for two reasons.  When we treat the offset cluster with a softened power law profile, the models require a large ($>30''$) core radius, which makes the ``cluster'' basically equivalent to a uniform mass sheet on the $\sim6''$ scale of strong lensing and thus negates the effect of having an extra halo.  NFW models cannot mimic that behavior, of course, since they do not have a flat core.  Instead, we find that an offset NFW cluster must have an unreasonably large ellipticity of $>0.7$.  We conclude that the strong lens data do not favor and in fact disfavor an offset cluster, so we not include a separate cluster component in our main analysis.

\subsubsection{Environmental terms}

While our lens-centered mass profiles should account for the majority of the mass in the strong lensing region (apart from a mass sheet), they should not be expected to represent the full complexity of the cluster potential.  The cluster halo may not be a simple ellipsoid on large scales, and it presumably has some lumpiness in the form of individual cluster galaxies.  In general, we can write the lens potential due to structures outside the strong lensing region using a Taylor series expansion of the form
\begin{eqnarray} \label{eq:cluspot}
\phi_{env}(r,\theta) &=& \frac{\kappa_c}{2}\,r^2 + \frac{\gamma}{2}\,r^2 \cos2(\theta- \theta_{\gamma}) \\
&& + \frac{\sigma}{4}\,r^3 \cos(\theta-\theta_{\sigma})  + \frac{\delta}{6}\,r^3 \cos3(\theta- \theta_{\delta}) + \ldots \nonumber
\end{eqnarray}
where $\kappa_c$ is a mass sheet, $\gamma$ is an external shear, and $\sigma$ and $\delta$ represent higher, third-order terms.  If the structures are ``far'' enough from the Einstein radius, the higher-order terms may be sufficiently small that the expansion can be truncated after the shear; indeed this approximation is used in many lens systems.  However, as shown by \citet{Kochanek91} and many subsequent studies, the third-order terms cannot be neglected in Q0957.  \citet{Bernstein99} and \citet{Keeton00} included the third-order terms but imposed the assumptions $\sigma = -2\delta/3$ and $\theta_\sigma = \theta_\delta$, which corresponds to the perturbation from a singular isothermal sphere.  We find that this assumption is too restrictive, so we include general third-order terms in order to allow more freedom in the models to account for complex features we have not explicitly modeled.

Note that we have written eq.~(\ref{eq:cluspot}) in terms of the amplitudes $(\gamma,\sigma,\delta)$ and directions $(\theta_\gamma,\theta_\sigma,\theta_\delta)$ of the shear and third-order terms.  We can think of these as ``polar'' coordinate versions of these parameters.  We can equivalently define ``Cartesian'' coordinate versions of the same parameters:
\begin{eqnarray}
  \gamma_c &=& \gamma\,\cos 2\theta_\gamma\,,  \hspace{0.5cm} \gamma_s = \gamma\,\sin 2\theta_\gamma\,, \nonumber\\
  \sigma_c &=& \sigma\,\cos  \theta_\sigma\,,  \hspace{0.637cm} \sigma_s = \sigma\,\sin  \theta_\sigma\,, \label{eq:cartparms} \\
  \delta_c &=& \delta\,\cos 3\theta_\delta\,, \hspace{0.59cm} \delta_s = \delta\,\sin 3\theta_\delta\,. \nonumber
\end{eqnarray}

While we quote results for the ``polar'' parameters, we actually carry out our modeling analysis using the ``Cartesian'' parameters.  The translation is straightforward.

In summary, our composite models consist of a stellar component, an elliptical halo centered on the stellar component that accounts for dark matter in the lens galaxy and/or cluster halo, and an additional set of terms corresponding to a third-order Taylor series expansion of the potential from the lens environment.  The 11 associated model parameters are listed in Table \ref{tab:parms}.

\subsection{Model Constraints}
\label{sec:constraints}

Table \ref{tab:data} lists the positions of the lensed images used as constraints on our models.  We supplement the faint images discussed in \S \ref{sec:features} with the quasar cores and radio jets resolved with Very Long Baseline Interferometry \citep[VLBI;][]{GorensteinVLBI,Garrett}.  With a resolution of $\sim0.1$ mas, the VLBI observations reveal structure within the jets and provide astrometry with formal errors $<1$ mas.  Previous studies of Q0957 used this excellent astrometry as strong constraints on lens models \citep{Barkana, Bernstein97, Bernstein99, Keeton00}.  In this work, however, we reconsider the use of such stringent constraints.  Many recent works have shown that lens galaxies often contain substructure in the form of CDM subhalos, which can perturb lens flux ratios by tens of percent or more and image positions at levels up to $\sim10$ mas \citep[e.g.,][]{Mao, Metcalf, Dalal, Chiba, Chen}.  Since our models do not contain any substructure,\footnote{It would certainly be interesting to add substructure to the models and use the precise radio positions and flux ratios to constrain the amount of substructure in the lens; but that is beyond the scope of the present work.} they should not be expected to match the image positions to better than the $\sim10$ mas residuals expected from substructure.  We therefore broaden the errorbars and adopt conservative uncertainties on the VLBI quasar and jet positions of 30 mas.  As a check, we verified that the VLBI derived positions of the quasars are in good agreement with the positions in our HST data.

We take the image position constraints together with constraints from our weak lensing analysis (\S \ref{sec:WL}) to comprise our main model constraints.  Subsequently, we consider adding supplementary constraints in the form of lensed arcs detected with HST/NICMOS, the quasar radio flux ratio, and stellar population synthesis models (see \S \ref{sec:results} for details).  Throughout the analysis we use the time delay $417.09 \pm 0.07$ days from \citet{Colley} to infer the Hubble constant.

\subsection{Strong Lensing Analysis}
\label{sec:SL}

In Bayesian language, our ultimate goal is to determine the posterior probability distribution for our model parameters and $H_0$, given the observational constraints.  We have $N_{data}=60$ position constraints, compared with $N_{param}=39$ free parameters (11 parameters for the mass model plus 28 source position parameters).  Hence our analysis has $N_{data}-N_{param}=21$ degrees of freedom.  The Hubble constant analysis involves one additional constraint (the observed time delay) and one additional parameter ($H_0$), so it has the same number of degrees of freedom.

Formally, the posterior probability distribution has the form
\begin{equation} \label{eq:Bayes}
  P(\theta|d,M)=\frac{P(d|\theta,M)\,P(\theta|M)}{P(d|M)}\ ,
\end{equation}
where $d$ denotes the data that provide constraints on the parameters $\theta$ associated with some model $M$.  The likelihood of the data given the model is
\begin{equation} \label{eq:like}
  \mathcal{L} \equiv P(d|\theta,M) \propto e^{-\chi^2/2}\ ,
\end{equation}
where $\chi^2$ is the goodness of fit.  The quantity $P(\theta|M)$ represents priors on the model parameters, which we take to be uniform.\footnote{Note that our uniform priors apply to the ``Cartesian'' coordinate versions of the environmental parameters (eq.~\ref{eq:cartparms}) and the ellipticity ($e_c = e \cos 2\theta_e$ and $e_s = e\sin 2\theta_e$).}  The Bayesian Evidence $P(d|M)$ is discussed below.

To ease the search of our large parameter space, we treat the image positions using the approximate position $\chi^2$ as defined by \citet{Keeton01}:
\begin{equation} \label{eq:chipos}
  \chi^2_p \ =\ \sum_i \delta\xv_i^T \cdot S_i^{-1} \cdot \delta\xv_i
  \ \approx\ \sum_i \delta\uv_i^T \cdot \mu_i^T \cdot S_i^{-1} \cdot \mu_i \cdot \delta\uv_i \,,
\end{equation}
where the sum runs over all images, $\delta\xv_i = \xv_{obs,i} - \xv_{mod,i}$ is the position residual for image $i$, $S_i = {\rm diag}(\sigma_i^2,\sigma_i^2)$ is the covariance matrix for the image positions, and $\mu_i$ is the lensing magnification tensor.  Using the lens equation, the position residual in the source plane is $\delta\uv_i = \xv_{obs,i}-\nabla \phi(\xv_{obs,i})-\uv_{mod}$.  The last step in eq.~(\ref{eq:chipos}) is valid when the position residuals are small such that the image and source plane residuals are related by $\delta\xv_i \approx \mu_i \cdot \delta\uv_i$.  The benefit of using this approach is that $\chi^2_p$ is quadratic in each source position $\uv_{mod}$, so we can find the best-fit value analytically:
\begin{eqnarray}
  \uv_{mod} = A^{-1} \bvec\,,
\end{eqnarray}
where
\begin{eqnarray}
  A &=& \sum_i \mu_i^T \cdot S_i^{-1} \cdot \mu_i \,, \\
  \bvec &=& \sum_i \mu_i^T \cdot S_i^{-1} \cdot \mu_i \cdot \uv_{obs,i} \,,
\end{eqnarray}
where these sums now run over the known images of a given source.  The upshot is that we do not have to search explicitly through the 28 dimensions corresponding to the source parameters.  An additional advantage is that we only have to compute the lens potential and its derivatives at the known positions of the images, which is useful for our dark matter models that require numerical integrals.

We must still search the 11-dimensional space of mass model parameters.  We sample the posterior probability distribution in this space using an adaptive Metropolis-Hastings Monte Carlo Markov Chain (MCMC) algorithm \citep{Roberts97, Haario, Roberts01}.  Each chain consists of a sequence of trial steps drawn from a multivariate Gaussian distribution.  In 95\% of the steps the Gaussian is based on an empirically updated covariance matrix to provide efficient sampling of a high-dimensional posterior distribution.  In the remaining 5\% of the steps the covariance matrix is diagonal so the algorithm takes independent and relatively large steps along the coordinate axes; this feature acts as a ``fail-safe'' to help the algorithm escape local minima in the $\chi^2$ surface and potentially discover new features in the posterior distribution.  The probability of accepting a trial step that modifies the likelihood from $\mathcal{L}_{old}$ to $\mathcal{L}_{new}$ is $\min(1,\mathcal{L}_{new}/\mathcal{L}_{old})$.  In other words, if a trial step increases the likelihood it is automatically accepted; while if it decrease the likelihood it is accepted with a probability given by the likelihood ratio.

We run 25 chains simultaneously and set them up to work ``from the outside in.''  That is, we generate an initial sample of models by drawing $\sim 10^4$ random points in the parameter space and optimizing them; this provides an estimate of the allowed parameter ranges (although without the proper statistical sampling that MCMC provides).  We then select models with maximal/minimal values of individual parameters to use as starting points for the MCMC chains.  By starting with well spread chains, the MCMC algorithm can establish the covariance matrix more quickly, and spend more time sampling the tails of the distribution, than it would by starting with closely-spaced starting points \citep{Gelman}.  The choice of initial models does not matter in detail, though, because for our final sampling we merge only the second halves of the chains in order to avoid sensitivity to the initial ``burn-in'' phase.

To assess whether an MCMC run has converged, we use the criterion presented by \citet{Gelman}.  For any given parameter $\theta$ we define
\begin{eqnarray}
  R(\theta) = \left[\frac{ {\rm var}(\theta) }{
    \frac{1}{J}\sum_{j=1}^{J}{\rm var}_j(\theta) }\right]^{1/2} ,
\end{eqnarray}
where ${\rm var}_j(\theta)$ is the variance of $\theta$ in the individual chain $j$, and ${\rm var}(\theta)$ is the variance of $\theta$ over the entire set of $J$ chains.  Heuristically, $R$ is a comparison of the variance of the entire distribution (the numerator) and the variance within individual chains (the denominator).  The ratio will be greater than 1 for disjoint chains, and it will decrease and asymptotically approach 1 as the chains converge.  \citet{Gelman} find that stopping an MCMC run once $R$ reaches values below 1.2 provides a sufficient description of the target distribution for most samplings.  To be conservative we run until $R \le 1.1$ for \emph{every} parameter.  We then repeat the entire MCMC analysis a total of three times to obtain robust sampling of the target distribution.

In general we let all 11 model parameters vary simultaneously.  The only exception is the scale radius in models with an NFW dark matter halo.  There is a strong covariance between $r_s$ and other parameters, which produces a narrow, curved ridge in the likelihood surface that is difficult for MCMC algorithms to sample efficiently.  To deal with this challenge, we discretely sample $r_s$ in logarithmic steps ranging from $0.1''$ to $1000''$.  We checked that the median values of parameters in the posterior distributions do not vary more than 5\% between steps.  To combine results from individual $r_s$ runs into the final posterior distribution, we need to normalize the individual results properly using eq.~(\ref{eq:Bayes}).  In particular, we need to determine the normalization factor
\begin{eqnarray}
  P(d|M)=\int P(d|\theta,M)\ P(\theta|M)\ d\theta\,,
\end{eqnarray}
which is known as the marginal likelihood or Bayesian Evidence.  Computing this integral usually requires techniques that are more computationally intensive than basic MCMC sampling, such as thermodynamic integration \citep[e.g.,][]{Marshall,Lartillot}.  However, it is possible to obtain a simple and useful surrogate for the evidence using the Bayesian Information Criterion,
\begin{eqnarray}
 {\rm BIC}=-2 \ln \mathcal{L}_{max} + k \ln N\,,
\end{eqnarray} 
where $\mathcal{L}_{max}$ is the maximum likelihood of the model, $k$ is the number of free parameters, and $N$ is the number points used in the fit.  While the BIC does not provide a proper statistical treatment of the evidence, it has been widely used in statistics and astronomy \citep[e.g.,][]{Schwarz, Rapetti, Davis, Liddle} and is sufficient for this study.

As discussed in \S \ref{sec:results}, we first examine models constrained only by the positions of the strongly lensed images, and then add supplemental constraints from weak lensing and various other considerations.  In practice, this means we run the MCMC analysis to sample the likelihood eq.~(\ref{eq:like}) based on the goodness of fit $\chi^2_p$ from eq.~(\ref{eq:chipos}).  (Since we use uniform priors, the posterior probability distribution is proportional to the likelihood.)  Now suppose we want to add some supplemental constraints characterized by their own goodness of fit $\chi^2_s$.  The total $\chi^2$ is just the sum $\chi^2_p+\chi^2_s$ (i.e., $\chi^2$ values add and likelihoods multiply), so we now want to sample
\begin{equation}
  P(\theta|d,M) \propto e^{-\chi^2_s/2} e^{-\chi^2_p/2} .
\end{equation}
Since the MCMC analysis provides a set of points drawn from $e^{-\chi^2_p/2}$, all we need to do is take those points and re-weight them by a factor proportional to $e^{-\chi^2_s/2}$.  This provides a simple way to apply additional constraints without having to repeat the full MCMC analysis.

\subsection{Weak Lensing Analysis}
\label{sec:WL}

In parallel with the strong lensing analysis, we have conducted a weak lensing analysis of the full $6'\times6'$ map to constrain the mass sheet and other environmental terms in the lensing potential (see eq.~\ref{eq:cluspot}).  The techniques and results of our weak lensing analysis are presented by \citet{Nakajima}. We find the cluster potential to be consistent with a core-softened isothermal sphere profile, $\kappa(r)=\kappa_0 [1+(r/r_c)^2]^{-1/2}$, with a best-fit central convergence $\kappa_0=0.47\pm0.17$ for a core radius $r_c=5.0''$, corresponding to a velocity dispersion of $(420\pm70) $ km s$^{-1}$ for $h=0.70$.   Additionally, we find the cluster to be consistent with an NFW profile but are unable to provide useful constraints on the cluster concentration and scale radius.

One product of our weak lensing analysis is the average convergence within $30''$ of the lens: $\kappa_{w,30''}=0.166\pm0.056$.  This represents the net convergence including contributions from both the lens galaxy and the cluster.  As discussed in \S \ref{sec:theory}, to determine the Hubble constant we need to know the correction from the cluster mass sheet,
\begin{eqnarray} \label{eq:sheetfac}
1-\kappa_c=\frac{1-\kappa_{w,30''}}{1-\kappa_{s,30''}}
\end{eqnarray}
where $\kappa_{w,30''}$ is the net convergence inside $30''$ from our weak lensing measurement while $\kappa_{s,30''}$ is the contribution from our strong lensing model.  Since we do not impose \textit{a priori} limits on $\kappa_{s,30''}$ in our models, the possibility exists that $\kappa_{s,30''}$ could exceed $\kappa_{w,30''}$.  In this sense, our measurement of $\kappa_{w,30''}$ provides an upper bound on  strong lens models.  We penalize models with $\kappa_{s,30''} > \kappa_{w,30''}$ by adding an additional $\chi^2$ term of the form
\begin{equation}
\chi^2_{\kappa} = \begin{cases}
\frac{(\kappa_{s,30''}-\kappa_{w,30''})^2}{\sigma^2_{\kappa_{w,30''}}}, & \kappa_{s,30''}>\kappa_{w,30''} \\
0, & \kappa_{s,30''}<\kappa_{w,30''}
\end{cases}
\end{equation}
When we apply the mass sheet correction to $H_0$ and mass model parameters, we need to account for the measurement uncertainties in $\kappa_{w,30''}$.  We do this using Monte Carlo techniques.  Specifically, we take each model from our MCMC runs and generate a distribution of values for $1-\kappa_c$ by drawing from a Gaussian distribution for $\kappa_{w,30''}$ set by the measurement and uncertainty from our weak lensing analysis.  (The factor of $\kappa_{s,30''}$ in eq.~\ref{eq:sheetfac} comes from the model itself.)  We then fold this distribution into our final results reported in \S\S \ref{sec:basic}--\ref{sec:NFW}.

The weak lensing analysis also yields constraints on the average shear in an annulus centered on the lens galaxy extending from $20''$ to $40''$: the two ``Cartesian'' shear components are $\gamma_c=-0.009\pm0.045$ and $\gamma_s=0.092\pm0.045$, or equivalently the two ``polar'' components are $\gamma=0.093\pm0.045$ and $\theta_{\gamma}=47.8\pm13.9$ deg.\footnote{Note that the uncertainties are likely to be non-Gaussian for the ``polar'' components.}  To compare this measurement to our models, we calculate the mean shear in the same annulus.  After multiplying our value of the mean shear by the $1-\kappa_c$ correction, we impose the weak lensing results as constraints on the lens models through additional $\chi^2$ terms.  These constraints penalize models with unusually small or large shears.

\section{Results}
\label{sec:results}

We first present results from lens models based on our new HST/ACS data (\S \ref{sec:basic}).  We then consider adding additional constraints from the quasar radio flux ratio (\S \ref{sec:fluxratio}), stellar population synthesis models (\S \ref{sec:SPS}), and physical properties of NFW halos (\S \ref{sec:NFW}).  All values we report for the parameters $(\ML, b_d, \gamma, \sigma, \delta)$, as well as the dimensionless Hubble constant $h = H_0/(100\  \Hunits)$, are corrected for the mass sheet through the factor $1-\kappa_c$ (including the associated uncertainties; see \S \ref{sec:WL}).

\subsection{Basic Results: Strong and Weak Lensing}
\label{sec:basic}

\begin{figure*}[!ht]
\centering
\includegraphics[width=12cm]{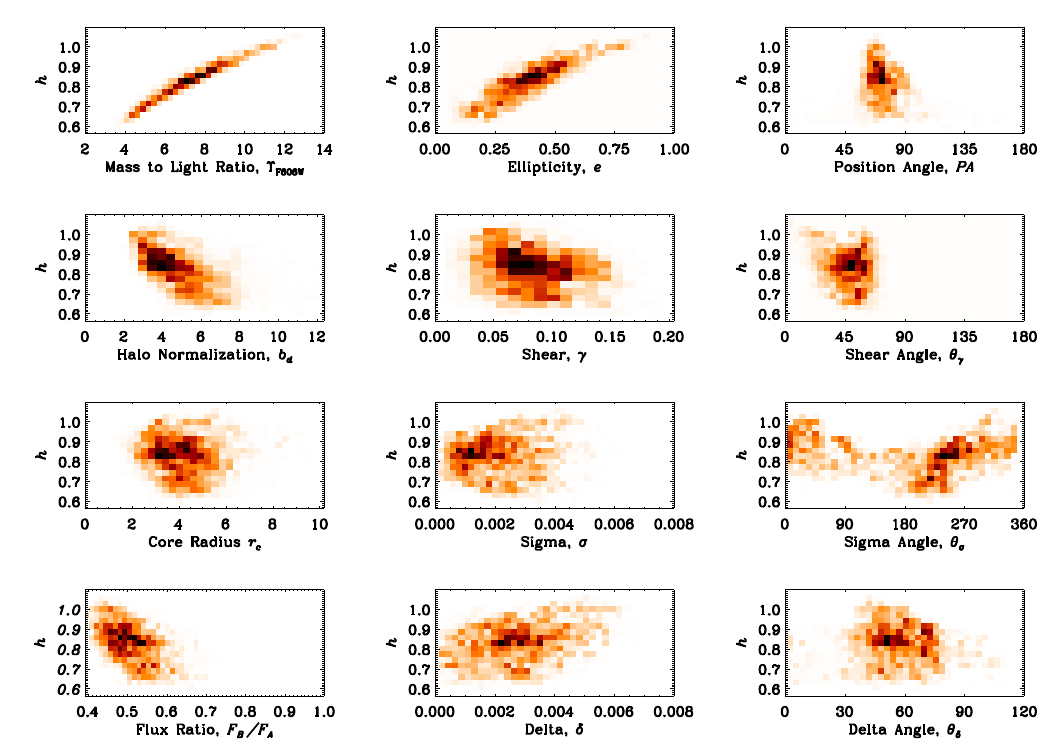}
\caption{2D histograms depicting the marginalized joint probability distributions $P(\theta,h)$ for each model parameter $\theta$ and the dimensionless Hubble constant $h$.  (For an explanation of parameters, see Table \ref{tab:parms}.)  We also show the model flux ratio in the lower left panel.  The colorscale is linear in the probability density, running from black at the peak to white at zero.  These results are based on dark matter models with a softened isothermal profile ($\alpha=1$).}
\label{fig:a102d}
\end{figure*}

\begin{figure*}
\centering
\includegraphics[width=12cm]{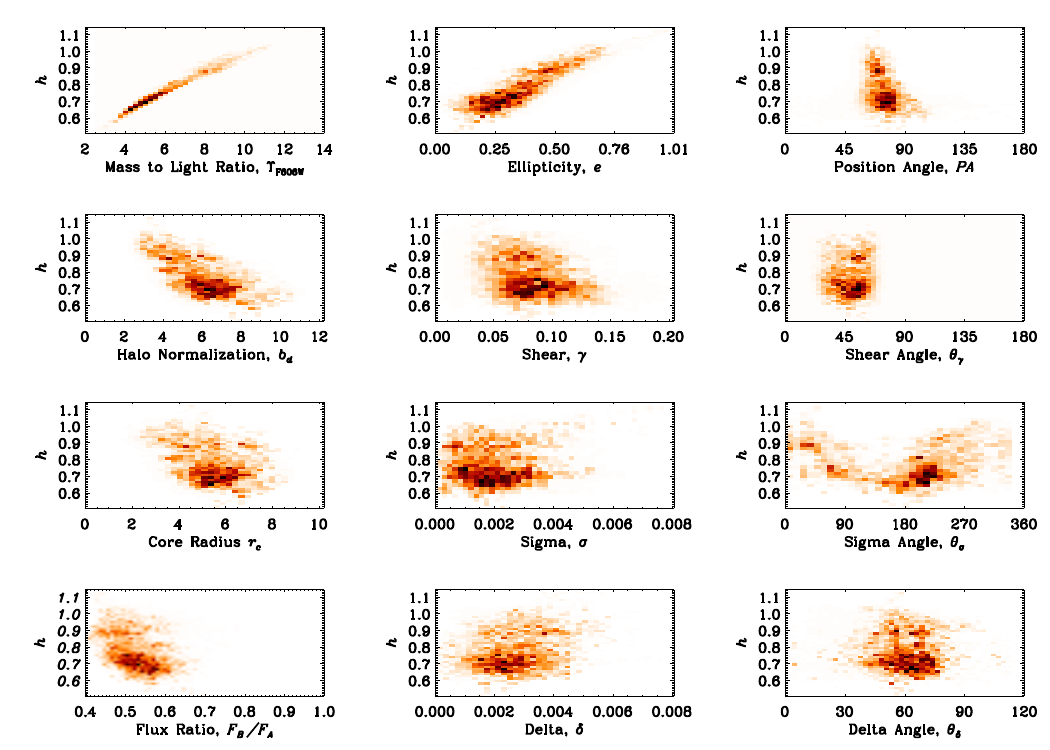}
\caption{Similar to Fig.~\ref{fig:a102d}, but for models in which the dark matter halo has a power law profile that is steeper than isothermal ($\alpha=0.5$).}
\label{fig:a052d}
\end{figure*}

\begin{figure*}[!ht]
\centering
\includegraphics[width=12cm]{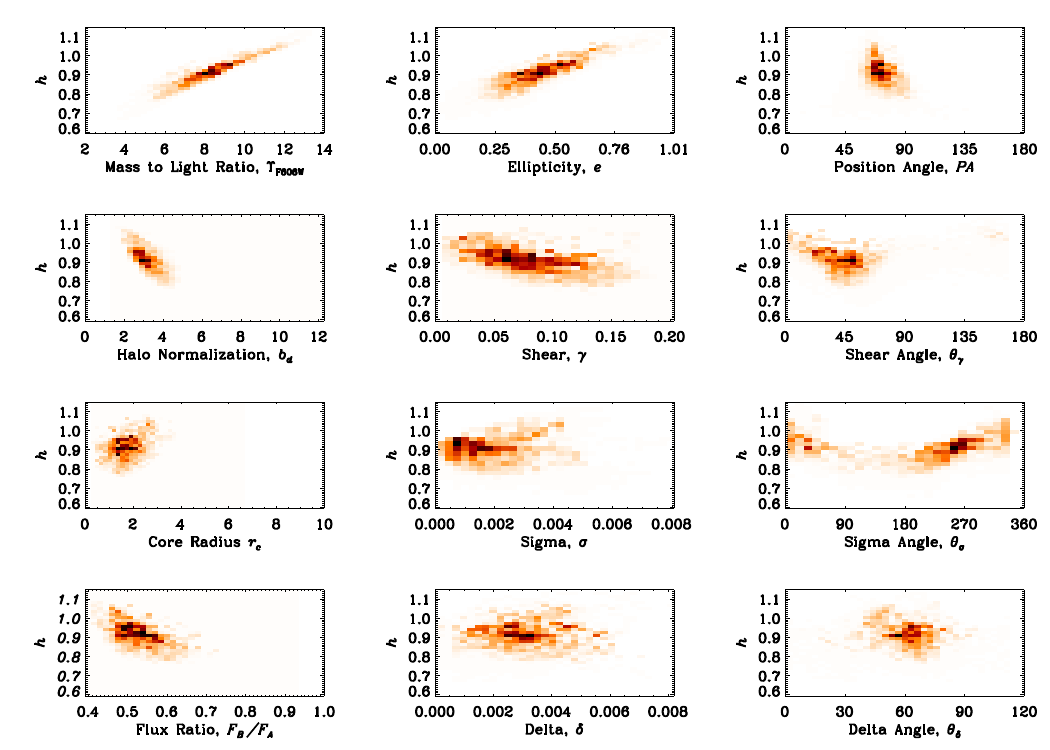}
\caption{Similar to Fig.~\ref{fig:a102d}, but for models in which the dark matter halos has a power law profile that is shallower than isothermal ($\alpha=1.5$).}
\label{fig:a152d}
\end{figure*}

\begin{figure*}
\centering
\includegraphics[width=12cm]{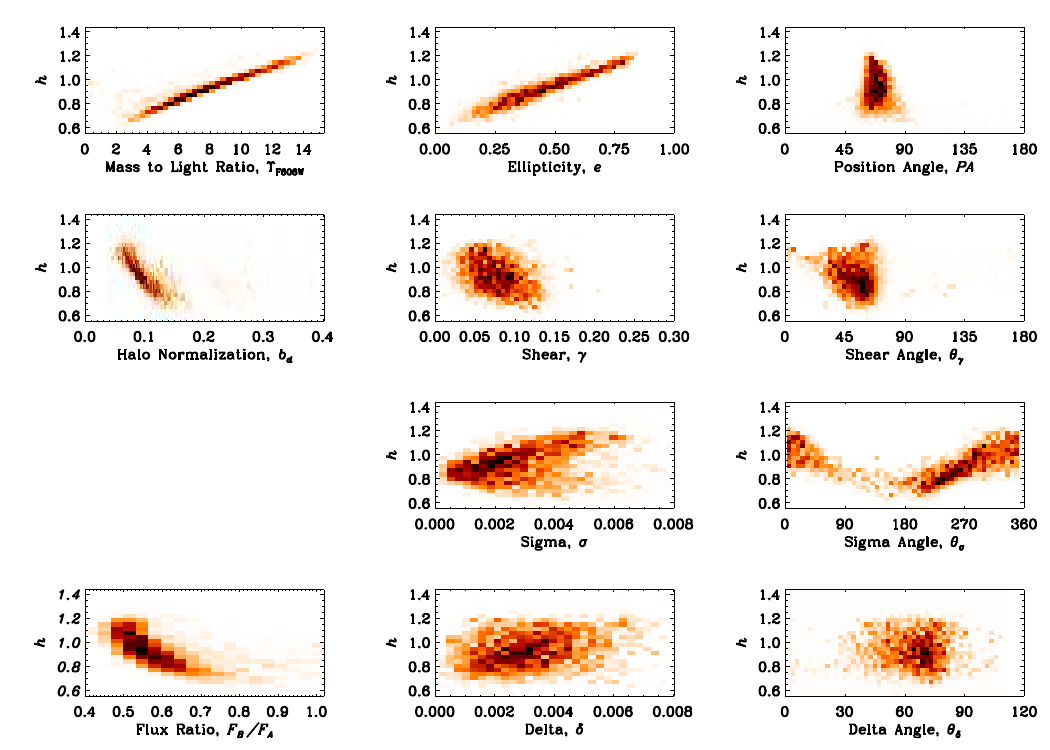}
\caption{Similar to Fig.~\ref{fig:a102d}, but for models with an NFW dark matter halo.  Here we do not show $P(r_s,h)$ since we discretely sample the scale radius (see \S \ref{sec:SL}).}
\label{fig:nfw2d}
\end{figure*}

For each of the four dark matter profiles we consider, we find a wide range of models that fit the HST strong lensing data well ($\chi^2_{reduced} < 1$).\footnote{As shown in Table \ref{tab:acsresults}, the best-fit model in each model class has a reduced $\chi^2$ somewhat less than unity.  We used the $\chi^2$ probability distribution to check that these values are statistically reasonable given the numbers of degrees of freedom.}  For example, the stellar mass-to-light ratio can be anywhere in the range $\ML \sim 4$--12, the ellipticity $e \sim 0.1$--0.7, and shear $\gamma \sim 0.04$--0.12.  Table \ref{tab:acsresults} lists the median value and 68\% and 95\% confidence intervals for each model parameter (from the individual marginalized posterior probability distributions).  The Table also lists the relative probabilities of our four dark matter models obtained by integrating the posterior probability distributions over $h$, after weighting by the BIC.  The range of allowed models is striking given that we now have so many strong lensing constraints.  Examining the relative probabilities, we find an isothermal ($\alpha=1$) dark matter profile is favored from lensing alone.  It is interesting to note, however, that our isothermal models require a values of $\kappa_{s,30''}$ which are larger than, but still consistent with our weak lensing measurements.

\begin{figure}
\centering
\includegraphics[clip=true, trim=1.4cm 0cm 0cm 0cm,width=9cm]{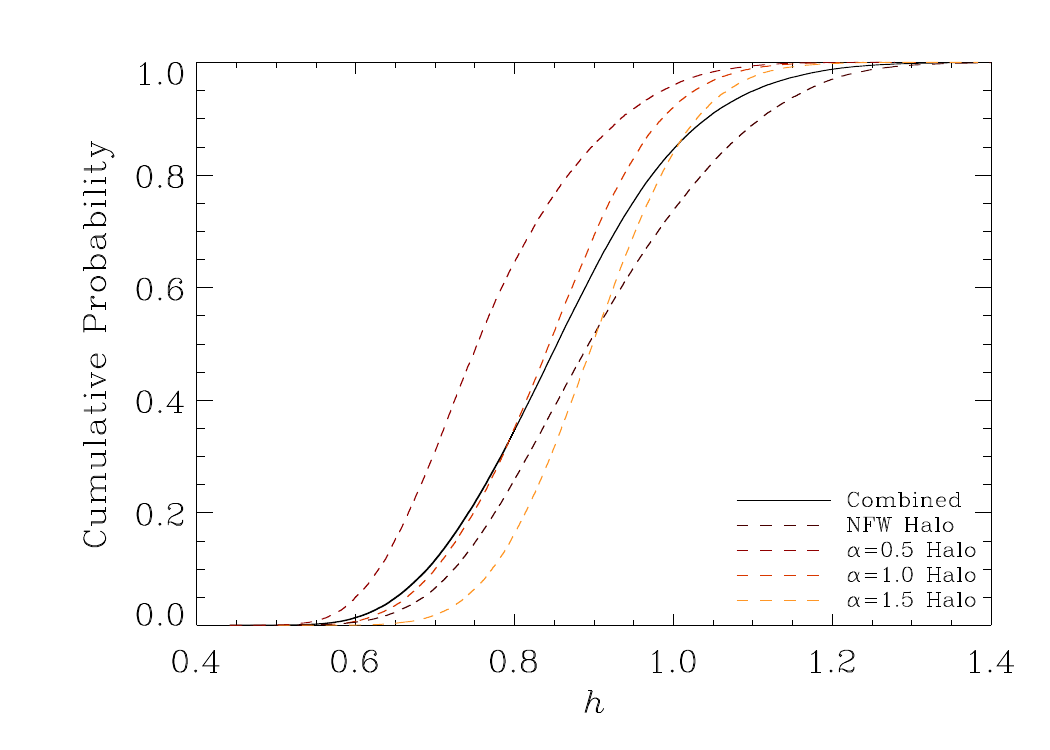}
\caption{Cumulative posterior probability distribution for the dimensionless Hubble constant $h$ for our four dark matter halo profiles.  The solid black curve shows the result of combining these distributions, weighted by their relative probabilities.}
\label{fig:hacs}
\end{figure}

There are various ways to examine the results, so let us begin with the Hubble constant.  Figures \ref{fig:a102d}--\ref{fig:nfw2d} show the marginalized joint probability distributions $P(\theta,h)$ for each model parameter $\theta$ and the dimensionless Hubble constant $h$, for all four classes of dark matter models.  Viewing the results this way helps reveal any important degeneracies or systematics that affect the inferred value of $h$.  The most obvious feature is a strong degeneracy between $h$ and the stellar mass-to-light ratio, $\ML$, which we discuss momentarily.  Focusing on $h$ itself, Figure \ref{fig:hacs} shows the marginalized cumulative posterior probability distribution for $h$ from each of our dark matter models.  Combining the four models (weighted by their relative probabilities), we find \hacs.  Our measurement of $H_0$ is somewhat higher than, but statistically consistent with, the recent determinations of $H_0=74.2\pm3.6$ \Hunits\ from SNe \citep{Riess} and $H_0=70.5\pm1.3$ \Hunits\ from WMAP5+BAO+SNe \citep{Komatsu}.  Compared with previous results from Q0957 \citep{Bernstein99, Keeton00}, our initial results have lowered the median\footnote{The old median value for $H_0$ is not very well determined, because the previous analyses did not include the proper statistical sampling provided by our MCMC analysis.} value from $\sim90$ to 85 and reduced the spread by $\sim 30\%$.  The latter result is significant given the new complexity in our models, the relaxation of (previously tight) quasar and jet positions, and the elimination of the quasar flux ratio constraint.

\begin{figure}[t]
\centering
\includegraphics[clip=true, trim=1.4cm 0cm 0cm 0cm,width=9cm]{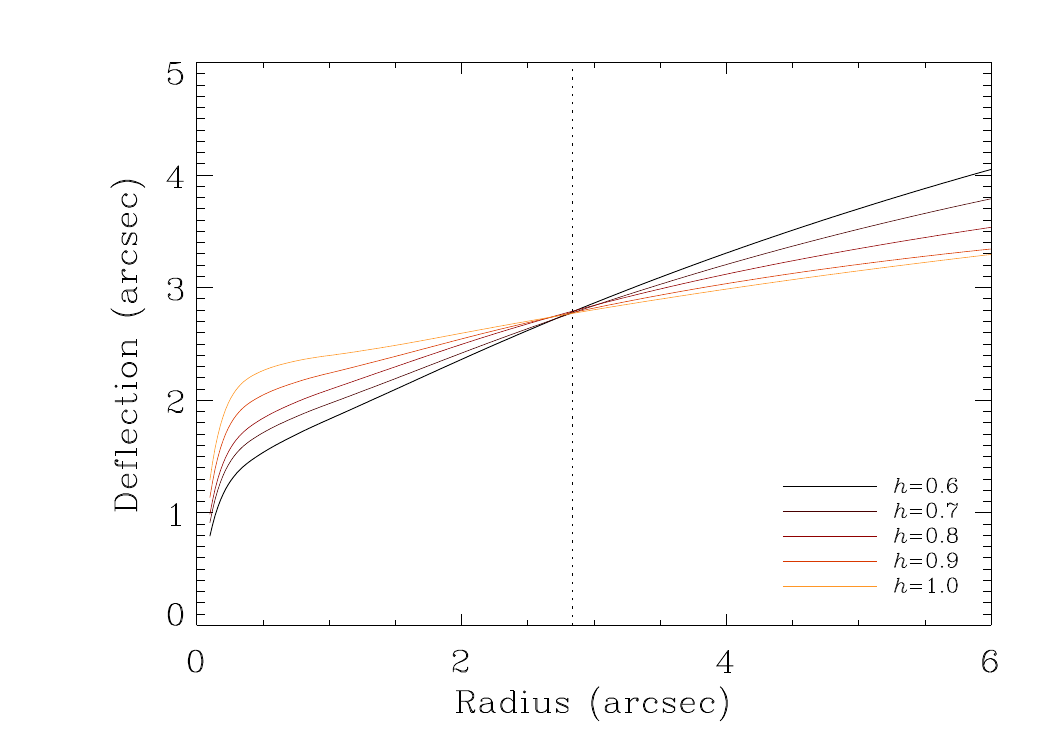}
\caption{Monopole deflection profile for models with $h=0.60$, 0.70, 0.80, 0.90 and 1.00.  We actually plot the mean profile for all models within $\pm0.005$ of the nominal $h$ value; the scatter among such models is $<2$\% across all radii and does not depend systematically on the particular dark matter profile.  The vertical dotted line indicates the Einstein radius at $\Rein=2.84''$.  The scatter among models drops to $<0.5$\% in the vicinity of $\Rein$, indicating a robust and tight constraint on the Einstein radius.}
\label{fig:defplot}
\end{figure}

The degeneracy between $h$ and the stellar mass-to-light ratio $\ML$ arises because the total mass within the Einstein radius is fixed, so varying $\ML$ changes the balance between the concentrated stellar component and the more diffuse dark matter halo.  That, in turn, modifies the slope of the total density profile, which is known to be the main factor that determines $h$ \citep[e.g.,][]{Williams00, Kochanek02}.  To illustrate these effects, we examine the monopole deflection curve, $\alpha(r) \propto M(r)/r$ where $M(r)$ is the projected mass within radius $r$ \citep[e.g.,][]{Blandford86,Blandford87,Cohn,Kochanek06,vandeven09}.  This is a 2D analog of the rotation curve.  A flat deflection curve corresponds to an isothermal profile, while a rising (falling) curve corresponds to a profile shallower (steeper) than isothermal. Figure \ref{fig:defplot} shows the deflection curves for models with different values of $h$.  There is a systematic change in the slope of the deflection curve with $h$, with very little ($<2$\%) scatter among models at a fixed value of $h$.  In other words, even though Q0957 has a complex lens potential, we still recover the familiar result that the slope of the total density profile is what principally determines $h$.  In our models, the slope of the density profile is governed by the stellar mass-to-light ratio, which makes $\ML$ the key parameter responsible for the range of $h$ values.  We consider external constraints on $\ML$ in \S \ref{sec:SPS} below.

One interesting aspect of Figure \ref{fig:defplot} is that Q0957 appears to have a rising deflection curve, corresponding to a total density profile that is shallower than isothermal, for any reasonable value of $h$.  Many other lens galaxies have profiles that are closer to isothermal \citep[e.g.,][]{Cohn,Rusin,Koopmans}.  Q0957 is not, however, unique in this regard: \citet{Kochanek06} found that HE 0435$-$1223 also has a rising deflection curve. They argue that different density profiles and deflection curves can arise as a consequence of how galaxies populate dark matter halos.  In the halo model (see \citealt{Cooray02} for a review), a group or cluster of galaxies consists of a massive central galaxy surrounded by smaller satellite galaxies.  Lying as it does at the center of the potential well, the central galaxy should have a higher dark matter surface density compared to its satellite neighbors, which would lead to a more diffuse mass distribution with a shallower profile and hence a rising deflection curve.  In this context, \citet{Kochanek06} argue that HE0435 may be the central galaxy in a group of galaxies.  Q0957 seems to fit naturally into this picture because it lies at or near the center of a modest cluster of galaxies (see \S \ref{sec:models}).

\begin{figure*}[t]
\centering
\includegraphics[clip=true, trim=2.5cm 0cm 4.5cm 0cm, width=7cm]{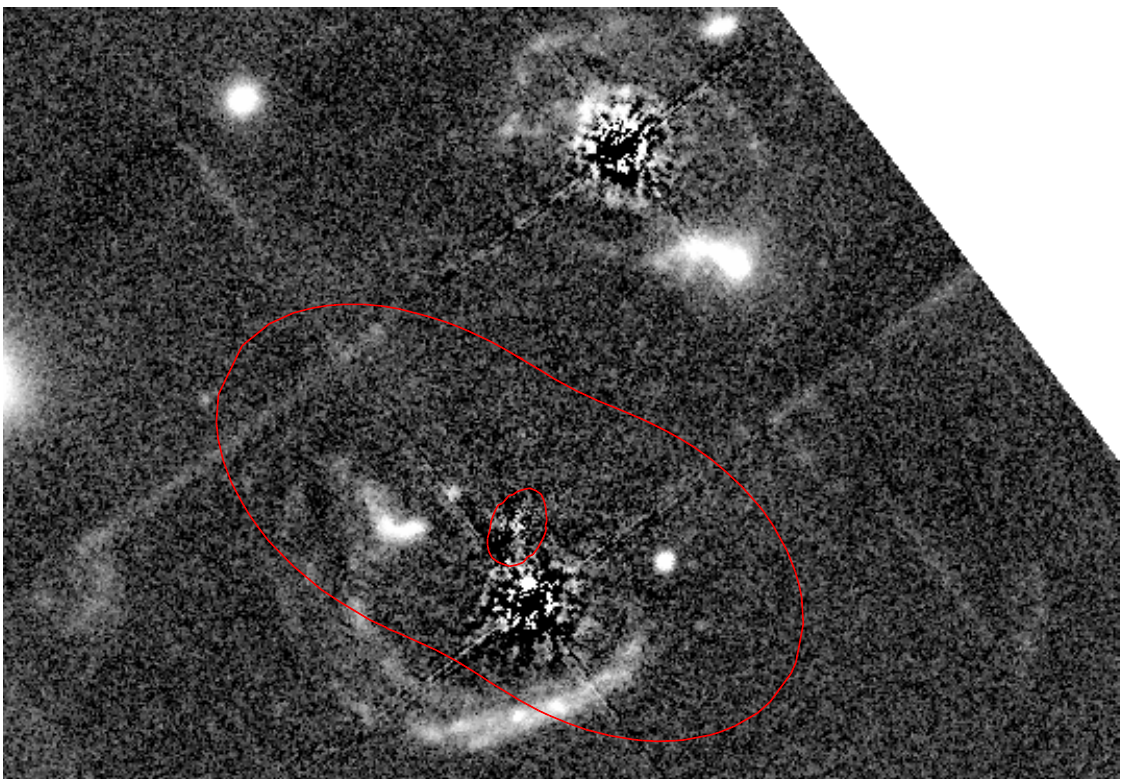}
\includegraphics[clip=true, trim=2.5cm 0cm 4.5cm 0cm, width=7cm]{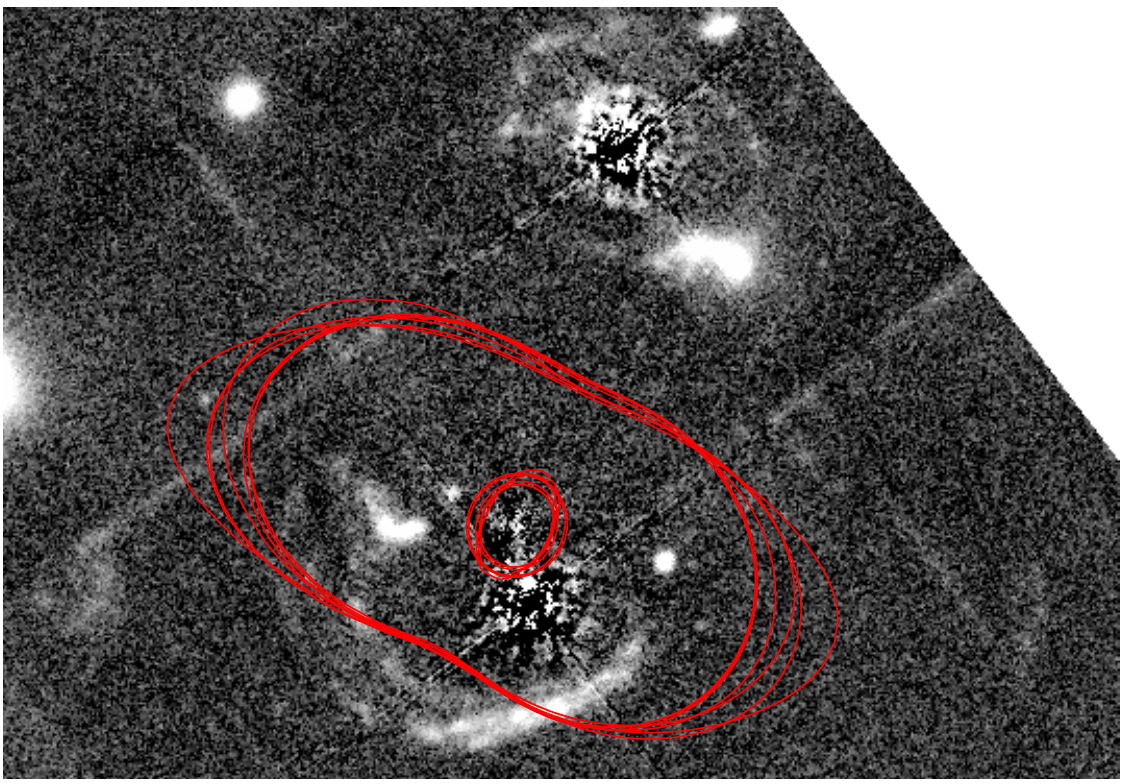}
\caption{
\textit{(a, left)} A typical critical curve resulting from the new image constraints in Table 2.  Notice the fold image pairs that the curve runs though South and East of quasar B.
\textit{(b, right)}  Critical curves corresponding to models with minimal/maximal values of different parameters, for our models with an NFW dark matter halo (results are similar for other model classes).  The critical curves show little variation along the semi-minor axis due to strong constraints from new images.  Significant variation still exists along the semi-major axis of the curves.}
\label{fig:crit}
\end{figure*}

One useful way to characterize a strong lens system is with the lensing critical curves and caustics.  The critical curves reveal highly magnified regions in the image plane, and the corresponding caustics separate regions in the source plane that lead to different numbers of lensed images.  Figure \ref{fig:crit} shows examples of the critical curves for our models of Q0957.  It is clear that the newly identified HST/ACS features have tightly constrained where the critical curve lies, especially southeast of quasar B. Significant variations do still exist near the ends of the critical curves, suggesting that constraints from the faint, unused features indicated in Figure \ref{fig:new} could help to further constrain the critical curves and restrict the parameter space.  Unfortunately, in the current data these features have a signal to noise ratio less then 3, making them not sufficiently reliable.

\begin{figure*}[b]
\centering
\includegraphics[width=8cm,height=8cm]{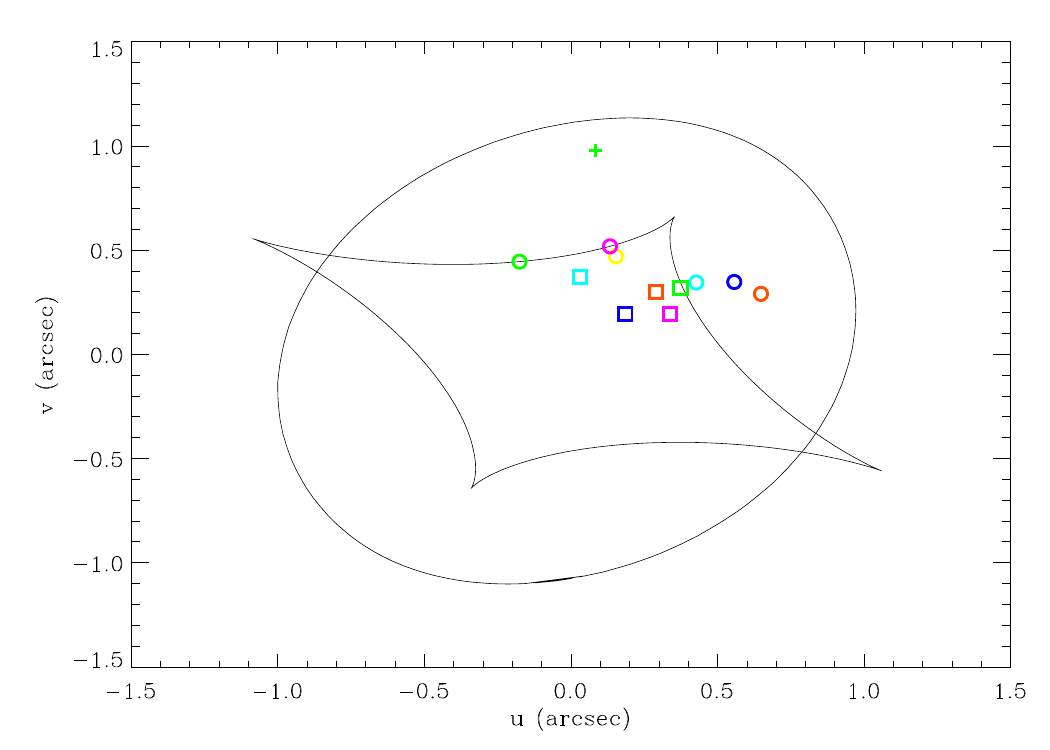}
\includegraphics[width=8cm,height=8cm]{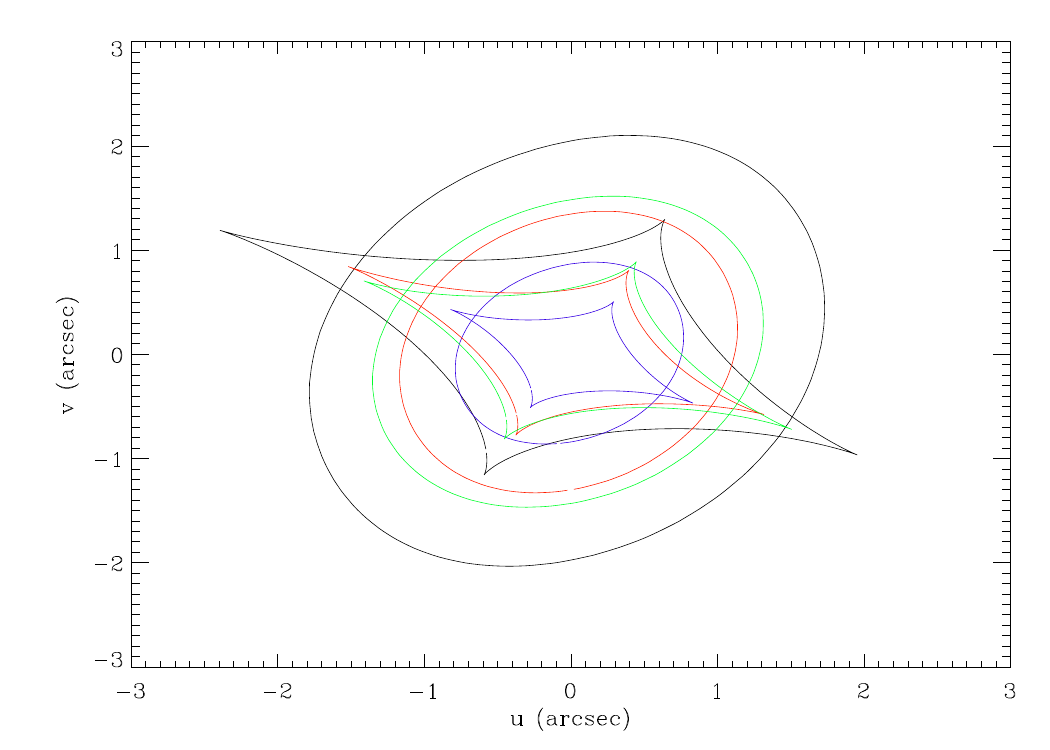}
\caption{
\textit{(a, left)} Source plane caustic corresponding to the critical curve shown in Fig.~\ref{fig:crit}a.  The points show the positions of the sources corresponding to the observed images, with the same shape and color scheme as in Fig.~\ref{fig:new}.
\textit{(b, right)} Source plane caustics corresponding to the critical curves shown in Fig.~\ref{fig:crit}b. The main variation is an overall rescaling due to different values of the convergence in the vicinity of the Einstein radius.}
\label{fig:caus}
\end{figure*}

In addition to $\ML$, our models exhibit a second degeneracy between $h$ and the ellipticity of the dark matter halo.  While the exact origin of this is unclear, it is likely connected to the degeneracy in $\ML$.  Since the stellar component has a fixed angular structure whose position angle ($\sim 67^\circ$--$82^\circ$, Fig.~\ref{fig:e-and-pa}) does not quite match the necessary angular structure of the lens potential (with position angle closer to $63^\circ$, see Fig.~\ref{fig:crit}), the dark matter component needs to make up the difference.  The amount of compensation increases as the mass of the stellar component increases, so the halo ellipticity rises with $\ML$.

\begin{figure*}[b]
\centering
\includegraphics[clip=true, trim=4cm 8cm 7cm 7cm, width=7cm]{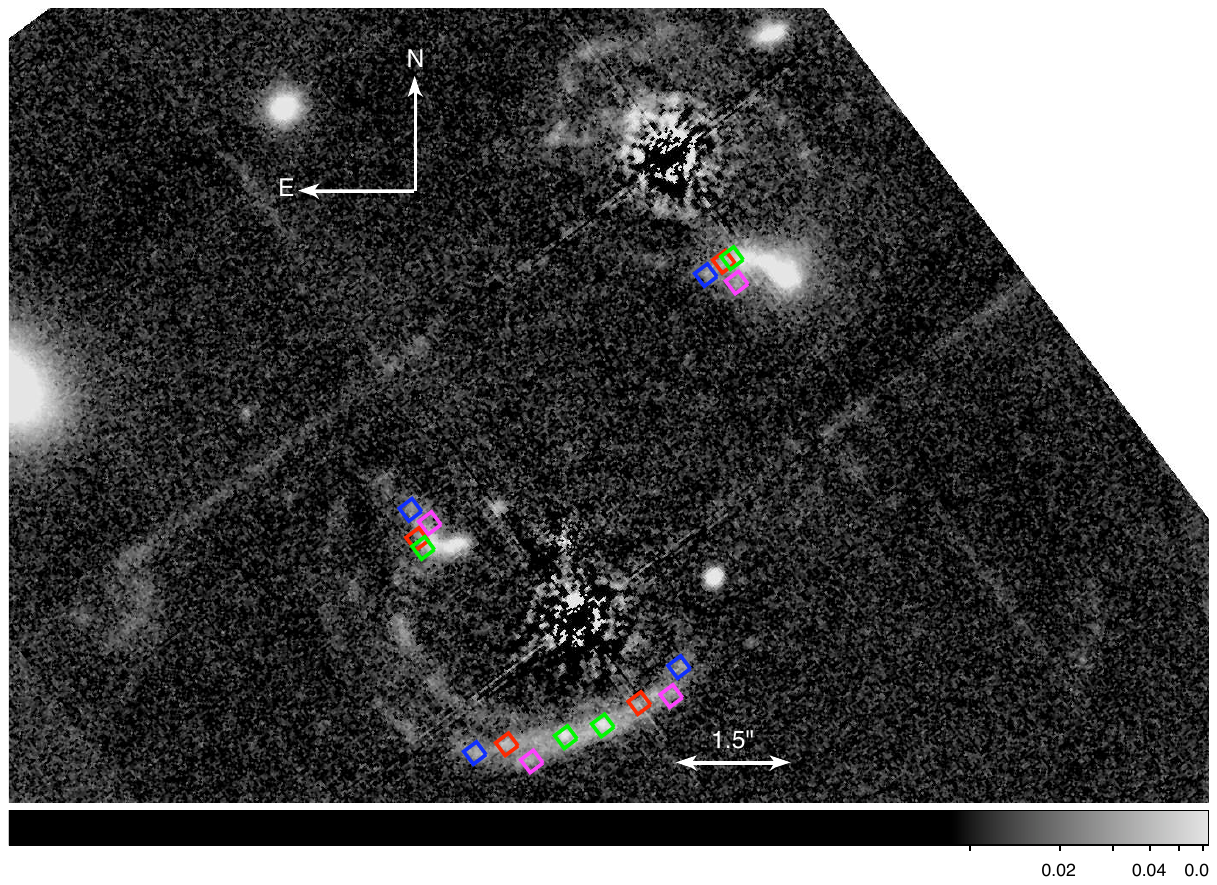}
\includegraphics[clip=true, trim=4cm 8cm 7cm 7cm, width=7cm]{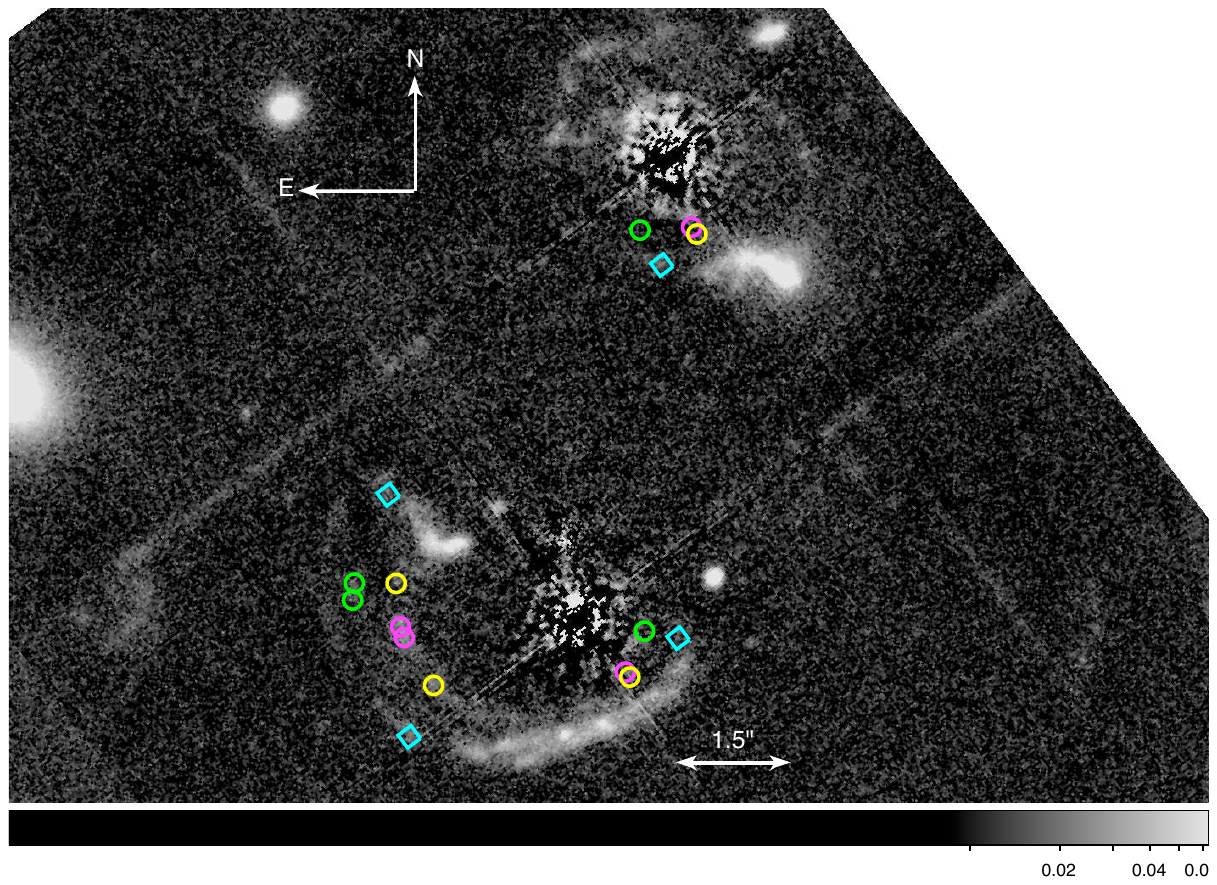}
\caption{
Predicted images of all sources that lie inside the tangential caustic for our lens models.  Results are shown for the same model as Figure \ref{fig:caus}a but are similar for other models.  The two panels correspond to different sets of sources, with the same arrangement as Figure \ref{fig:new}.  There are some predicted images shown here that have not been detected (i.e., they do not appear in Fig.\ \ref{fig:new}); they are predicted to be below the noise in our HST data.}
\label{fig:fourimage}
\end{figure*}

Figure \ref{fig:caus} shows the corresponding caustics in the source plane. Generically, the tangential caustic is elongated along a roughly NE-SW direction and extends beyond the radial caustic.  There is some variation among the models, but the main effect is just an overall rescaling by $1-\kappa_E$, where $\kappa_E$ is the convergence at the Einstein radius (which is related to $\kappa_{s,30''}$).  Many of our newly discovered sources are inferred to lie within the tangential caustic and should therefore have four images.  Since we did not necessarily identify these as quad systems in our original detections (\S \ref{sec:features}), it is interesting to examine the predicted counter-images.  Figure \ref{fig:fourimage} shows all the predicted images of the quadruply-imaged sources, for comparison with the detected images shown in Figure \ref{fig:new}.  (We show results for one particular model, but results for other models are similar.)  We see that there are some predicted images that lie in relatively blank regions West of quasar B and South of quasar A.  This is not a concern, however, because the undetected images have magnifications that are a factor of 10--100 smaller than the detected images lying to the East of quasar B, so their predicted fluxes lie well below the noise level of the HST image.  Some of the predicted counter-images (among the ones indicated by diamonds) are not so far below the noise, but they still lack clear peak positions and thus cannot currently provide further robust constraints on lens models.

We consider one additional source of strong lensing constraints, namely extended images of the quasar host galaxy observed with HST/NICMOS by \citet{Keeton00}.  Following \citeauthor{Keeton00}, we analyze the arcs by taking the resolved arc around quasar A, mapping it pixel-by-pixel to the source plane using the lens model of interest, then mapping the reconstructed source back to the image plane to predict the arc around quasar B.  Generally, we find that all models generated from our ACS data reproduce the NICMOS arcs comparably well; the NICMOS arcs do not restrict the range of models significantly better than the ACS data.  We infer that the ACS data have captured most of the information present in the NICMOS arcs, which is not surprising given that the ACS and NICMOS data span similar spatial regions and (presumably) both come from the lensed quasar host galaxy.  Compared with the NICMOS arcs, the ACS data are somewhat cleaner to interpret because they avoid complicated interpolations to and from the image plane and offer a more straightforward counting of degrees of freedom.  We take the compatibility of the NICMOS and ACS data as additional reassurance that the mapping of faint lensed features (\S \ref{sec:features}) has been done correctly.

\subsection{Quasar Flux Ratio}
\label{sec:fluxratio}

So far we have only considered image positions as lensing constraints.  In order to include some information about the lensing magnification, we consider measurements of the quasar flux ratio.  \citet{Bernstein99} and \citet{Keeton00} used the delay corrected VLA measurements at 4 cm and 6 cm by \citet{Haarsma} to constrain the quasar core flux ratio.  While the VLA cannot resolve out the relative contributions of radio compontents, if the jets are invariant on decadal time scales, the ratio of the radio \textit{fluctuations} gives a measurement of the core flux ratio.  With this assumption, \citeauthor{Haarsma} derive the core flux ratio to be $0.74\pm0.02$.  Before applying this constraint to our lens models, we must consider whether our models should be expected to fit the observed flux ratio.  The issue is whether dark matter substructure may perturb the flux ratio \citep[e.g.,][]{Mao, Metcalf, Dalal, Chiba, Macleod} in a way that our smooth models cannot reproduce.

High-resolution VLBI measurements show that the quasar images are $\lesssim1.2$ mas in size \citep{GorensteinVLBI}, which corresponds to a size for the emission region in the source plane of $\lesssim 0.9$ mas or $\lesssim 5.4\,h^{-1}$ pc.  We use the methods of \citet{Dobler} to estimate how a source of this size would be affected by an isothermal sphere clump placed near one of the images.  We find that a clump of mass $\gtrsim 10^6\,M_\Sun$ can easily change the lensing magnification by a factor of order unity, and $N$-body simulations predict such subhalos to be abundant ($\gtrsim10^3$) in a galaxy with a mass of $\sim10^{13}\,M_{\Sun}$ \citep[e.g.,][]{Springel, Angulo}.  Apparently we should not discount the possibility that substructure plays a significant role in the observed VLA flux ratio.

Our basic models generally predict a flux ratio in the vicinity of $\sim0.5$ with at most a tail extended to the range of the VLA measurement (see Figs.\ \ref{fig:a102d}--\ref{fig:nfw2d} and Table \ref{tab:acsresults}).  The discrepancy could be interpreted as evidence that the VLA flux ratio is indeed perturbed by substructure.  Further support for this hypothesis comes from the fact that the magnification ratio inferred from the resolved radio jets is different from the ratio for the quasar cores, and closer to the smooth model prediction \citep{Bonometti, GorensteinVLBI, Conner}.  We should be careful, of course, not to think that a measurement that disagrees with our smooth models is ``wrong'' and one that agrees is ``right'' --- or to assume that any discrepancy involving a flux ratio automatically implies dark matter substructure.  Nevertheless, we conclude that existing evidence shows the flux ratio to be very intriguing and worthy of further study, both on its own and as possible evidence for substructure in Q0957.

\begin{figure}[t]
\centering
\includegraphics[clip=true, trim=1.4cm 0cm 0cm 0cm,width=9cm]{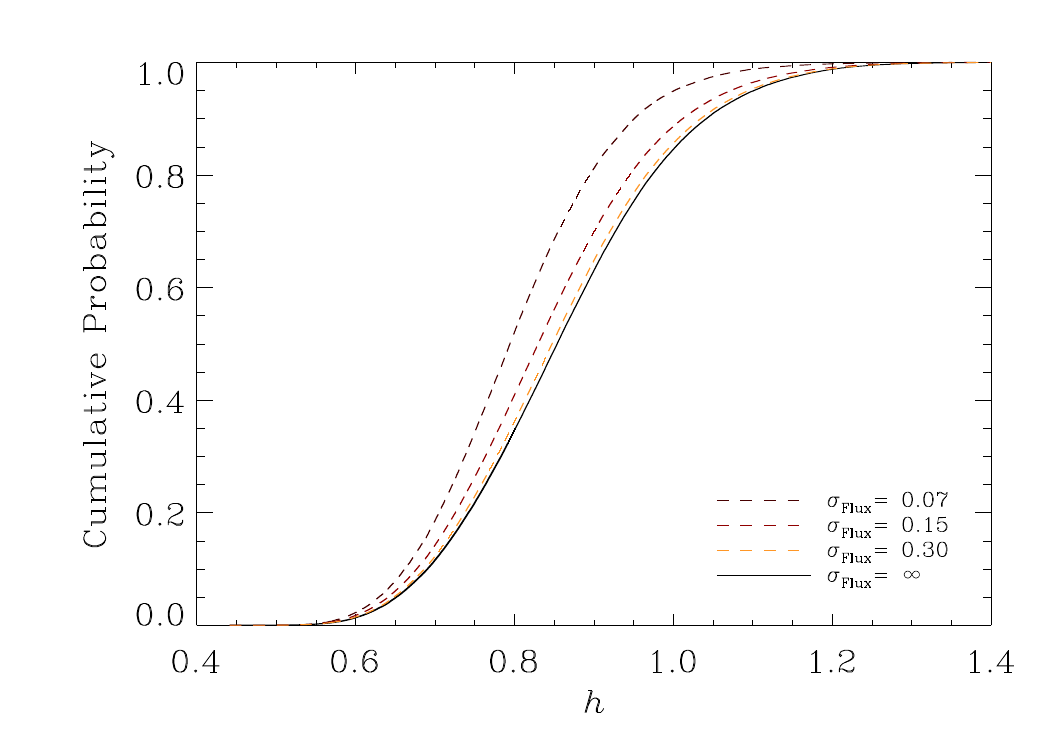}
\caption{Cumulative posterior probability distribution for the dimensionless Hubble constant $h$ with different assumptions about the constraint from the quasar radio flux ratio.  The solid black line shows our fiducial results using the image positions alone (from Fig.~\ref{fig:hacs}).  The dashed lines show how the results change when we impose constraints from the VLA flux ratio constraints \citep{Haarsma} with the uncertainties increased by a factor of 3.5, 7.5, and 15, representing different levels at which effects from substructure might be understood.  These values are consistent with a increase in the magnification of image B by a factor of 1.35, 1.20, and 1.0 due to a subhalo of mass $\gtrsim10^6 M_{\Sun}$.  While adding flux constraints clearly tightens the $h$ distribution, we do not include them in our final $h$ distribution since we do not currently understand the extent to which substructure may be important in Q0957.}
\label{fig:flux}
\end{figure}

With this in mind, it is not clear how strongly we should impose the VLA flux ratio as a constraint on our models.  We consider using the measurement but inflating the errorbar by various factors to obtain a range of constraints from strong to weak.  Figure \ref{fig:flux} shows the marginalized cumulative posterior probability distribution for $h$ for the different cases.  The flux ratio constraint tends to reduce the median value of $h$ and tighten the distribution.  While such a result seems enticing, we caution that it may be artificial if the flux ratio is really perturbed by substructure which is absent from our model.  Given the concerns, we choose not to use the flux ratio as a constraint for our final results.  If there were some way to determine the ``macro'' flux ratio, however, that might help improve constraints on $h$.

\subsection{Stellar Mass to Light Ratio}
\label{sec:SPS}

As shown in Figures \ref{fig:a102d}--\ref{fig:nfw2d}, our models demonstrate a strong correlation between $h$ and $\ML$, the stellar mass-to-light ratio, as a consequence of the radial profile degeneracy in lensing.  Any external constraints on $\ML$ could not only narrow the overall parameter space but also tighten constraints on $h$.  It is possible to obtain such constraints from stellar population synthesis (SPS) models thanks to the fact that we have excellent HST photometry of a relatively ``simple,'' old stellar population in the lens galaxy.

We use two sets of SPS models.  The first are the Flexible Stellar Population Synthesis (FSPS) models of \citet{Conroy}.  These models exhibit enormous flexibility and are aimed at addressing many of the uncertainties of SPS models.  In particular, the user can not only consider the traditional effects of varying the star formation history, star formation epoch, metallicity, initial mass function (IMF), and dust, but also account for various treatments of the thermally pulsating asymptotic giant branch (TP-AGB), blue stragglers, and blue horizontal branch stars.  See \citet{Conroy} for details of the FSPS models and the SPS uncertainties they address.  Using the FSPS models, we follow the treatment of \citeauthor{Conroy}, adjusting 9 models parameters (see Table \ref{tab:spsparms}) that include the effects of varying the epoch of star formation, the star formation history, and dust.  We consider six metallicities from 50\% to 160\% Solar using a Chabrier IMF \citep{Chabrier}.

The second set of SPS models we use are from \citet{Maraston08}.  These models are based on the same simple stellar populations (SSPs) as \citet{Maraston05}, which treat the TP-AGB contribution to the SEDs, but also include a metal poor ($[Z/H]=-2.2$) population comprising 3\% of the mass.  As shown by \citet{Maraston08}, these models provide a good fit to the optical colors of galaxies in the Luminous Red Galaxy (LRG) sample from the Sloan Digital Sky Survey.  This is encouraging because SPS models have historically had trouble fitting LRG colors \citep[e.g.,][]{Eisenstein, Wake}.  Since the lens galaxy in Q0957 is luminous ($L\sim6.5L_{\star}$) and red ($m_{F606W}-m_{F814W}=1.057$), we postulate that these models should provide a good fit the galaxy's spectral energy distribution.   To account for variation of the lens galaxy from the SSPs of \citeauthor{Maraston08}, we allow for variation in the redshift at which star formation begins, the star formation history, and dust.  As in our FSPS models, we allow star formation to begin anywhere from the CMB redshift, $z=1089$, to the redshift of the lens, $z=0.361$.  We adopt the star formation rate
\begin{eqnarray}
\Psi(t)&=&\frac{1-C}{\tau}\frac{e^{-t/\tau}}{e^{-T(z_{form})/\tau}-e^{-T(z_l)/\tau}} \\ &&+\frac{C}{T(z_l)-T(z_{form})},    \hspace{1.1cm}T(z_{form}) \le t \le T(z_l) \nonumber,
\end{eqnarray}
where $C$ is the fraction of stars formed at a constant rate, $\tau$ is the e-folding time of the star formation rate, and $T(z)$ denotes the age of the universe at redshift $z$.  This form of the star formation rate has the advantage of smoothly varying from a SSP to a constant star formation history.  For dust, we consider the two-parameter model of \citet{Charlot} which includes the effect of dust around young stars as well a diffuse dust component.  Parameters for our Maraston SPS models are summarized in Table \ref{tab:spsparms}.

As constraints on the SPS models, we use our measurements of the F606W and F814W magnitudes of the lens galaxy (see Table \ref{tab:galphot}) together with a reanalysis of the NICMOS/F160W image obtained by \citet{Keeton00}.  The revised AB magnitude $m_{\rm F160W,AB}=16.84\pm0.15$ (Chien Peng, private communication) differs somewhat from the originally reported value because it is based on a more sophistocated deconvolution of the lens galaxy from the quasar images and host arcs.   We correct for Galactic extinction using the methods of \citet{Schlegel}, finding the value of $E(B-V) = 0.0095$.

Initial modeling found that large values of dust extinction, extending to $A_{\rm F606W}>2.0$ mag, were allowed in the SPS models.  This was unexpected since early type galaxies are known to have modest dust content \citep[e.g.,][]{Schawinski}.  In lensed systems, differential extinction measurements have shown that lenses typically exhibit smaller values of dust extinction than found in nearby, late-type galaxies: \citet{Eliasdottir} find a mean extinction of $A_V=0.56\pm0.04$ in a sample of 10 lenses \citep[also see][]{Falco99}.  For Q0957 we could in principle rely on previous attempts to measure the dust content of the lens galaxy, but the results are puzzling.  \citet{Goicoechea05} used HST/STIS observations to measure the flux ratio $F_B/F_A>1$ at optical and ultraviolet wavelengths \citep[also see the delay corrected ratios of][]{Colley}, which stands in stark contrast to the VLA measurement $F_B/F_A=0.74\pm0.02$ and our models predictions.  To explain this difference \citeauthor{Goicoechea05} invoke dust clouds in front of image A leading to extinction $A_V=0.30$.  It is counterintuitive to think that image A (at $18.6\,h^{-1}$ kpc from the center of the galaxy) is more heavily extinguished than image B (just $3.7\,h^{-1}$ kpc from the center).  If extinction is indeed the cause of the wavelength dependence in the flux ratios, it remains unclear how the dust is distributed throughout the rest of the galaxy, whether it is clumpy and extends to large radii in other directions.  For all these reasons, we choose not to constrain the dust in the SPS models to a particular value.  Nevertheless, in order to avoid unreasonable large extinction values we impose a weak, exponential prior of the form $e^{-A_{F606W}/1.0 \rm{mag}}$.

\begin{figure*}[ht]
\centering
\includegraphics[clip=true, trim=1.4cm 0cm 0.2cm 0cm,width=8.75cm]{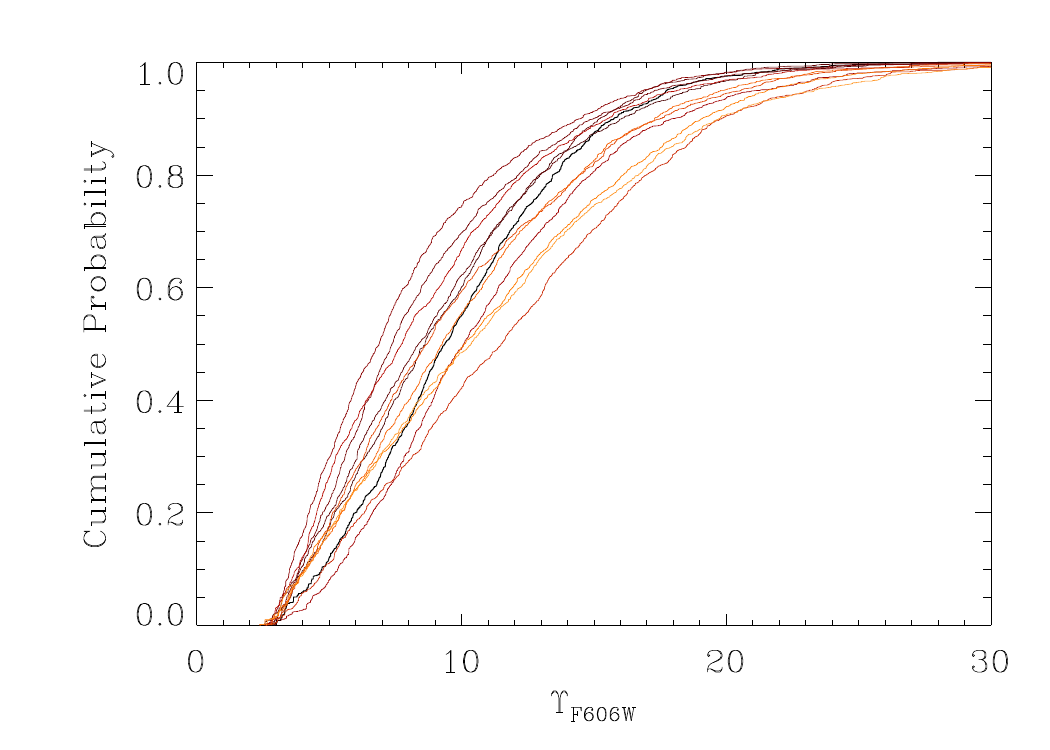} \hspace{0.3cm}
\includegraphics[clip=true, trim=1.4cm 0cm 0.2cm 0cm,width=8.75cm]{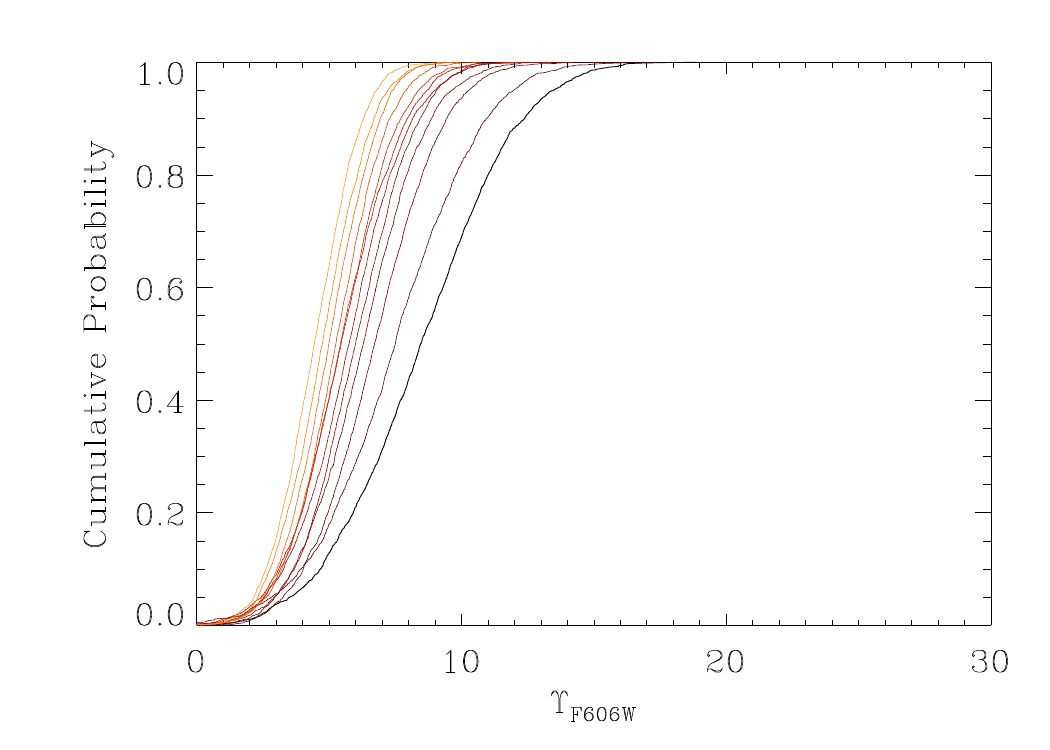}
\caption{Cumulative posterior probability distributions for the stellar mass to light ratio, $\ML$.  The different curves correspond to different values of $h$, varying from 0.55 (black) to 1.10 (light orange) in increments of 0.05.  \textit{(a, left)} Results from the FSPS models of \citet{Conroy}. \textit{(b, right)} Results from the SPS models of \citet{Maraston08}.  The FSPS models tend to produce higher values of $\ML$ with more scatter than the Maraston models.  Also, the FSPS models are not as \textit{systematically} dependent on $h$, presumably because the large freedom in the models dominates the distribution of $\ML$.}
\label{fig:ML}
\end{figure*}

In order to derive constraints on $\ML$, we set up an MCMC analysis similar to what we use for lens models (see \S \ref{sec:SL}) to sample the posterior probability distribution.  Since the value of $\ML$ depends on $h$ (principally through the age of the universe as a function of redshift), we run the MCMC analysis for discrete values of $h$ from 0.50 to 1.45 in steps of $\Delta h=0.025$; the small steps ensure that the median and range of $\ML$ do not vary by more than 3\% from one $h$ value to the next, so we can interpolate accurately.  Figure \ref{fig:ML} shows the cumulative distributions for $\ML$ for different values of $h$, from both FSPS and Maraston models.  In general, both models fit the observed F606W$-$F814W and F606W$-$F160W colors well ($\chi^2_{reduced}\le 1$), but the FSPS models yield a larger range for $\ML$.  This is not surprising given the amount of freedom available in the FSPS models.  Previous studies of massive ellipiticals found values of $\Upsilon_B$ of $\sim 4-10$ \citep[e.g.,][]{Gerhard,Grillo}, in good agreement with the values found here.\footnote{At a redshift of $z=0.36$, observed $\ML$ corresponds to rest frame $\Upsilon_B$.}

\begin{figure}[hb]
\centering
\includegraphics[clip=true, trim=1.4cm 0cm 0cm 0cm,width=9cm]{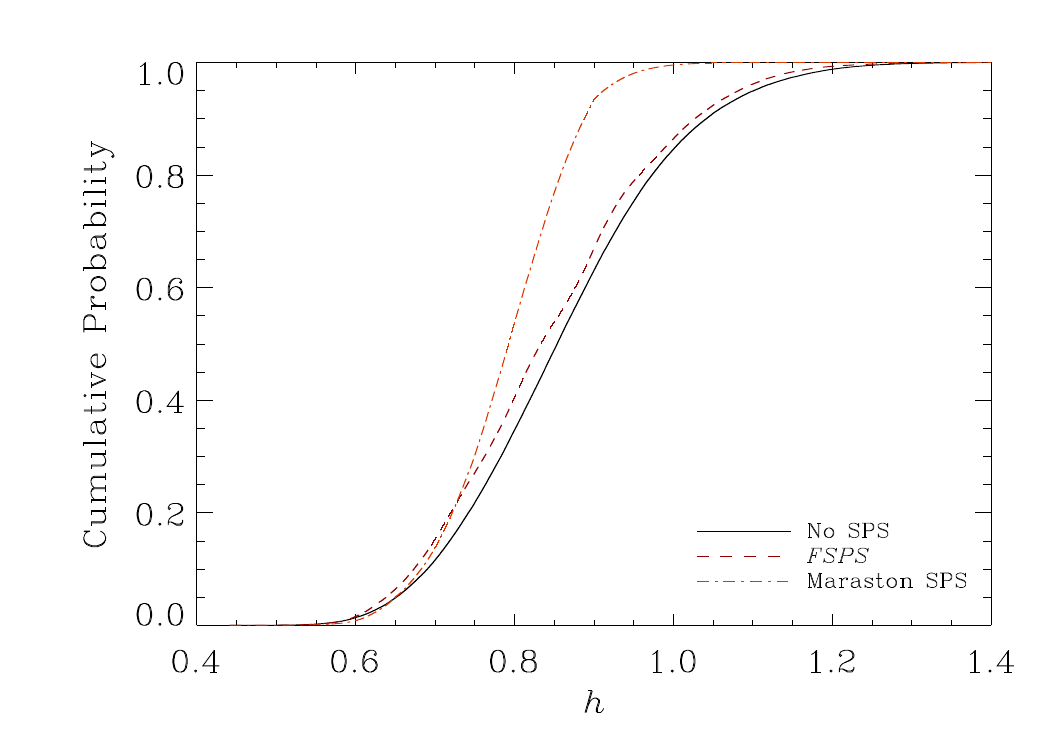}
\caption{
Cumulative posterior probability distribution for $h$ with and without constraints from SPS models.  Since the Maraston SPS models have been shown to fit LRG colors from the SDSS, we adopt them when quoting final values of $h$ and model parameters.  We find \hml.}
\label{fig:h_ml}
\end{figure}

\begin{figure}[b]
\centering
\includegraphics[clip=true, trim=1.4cm 0cm 0cm 0cm,width=9cm]{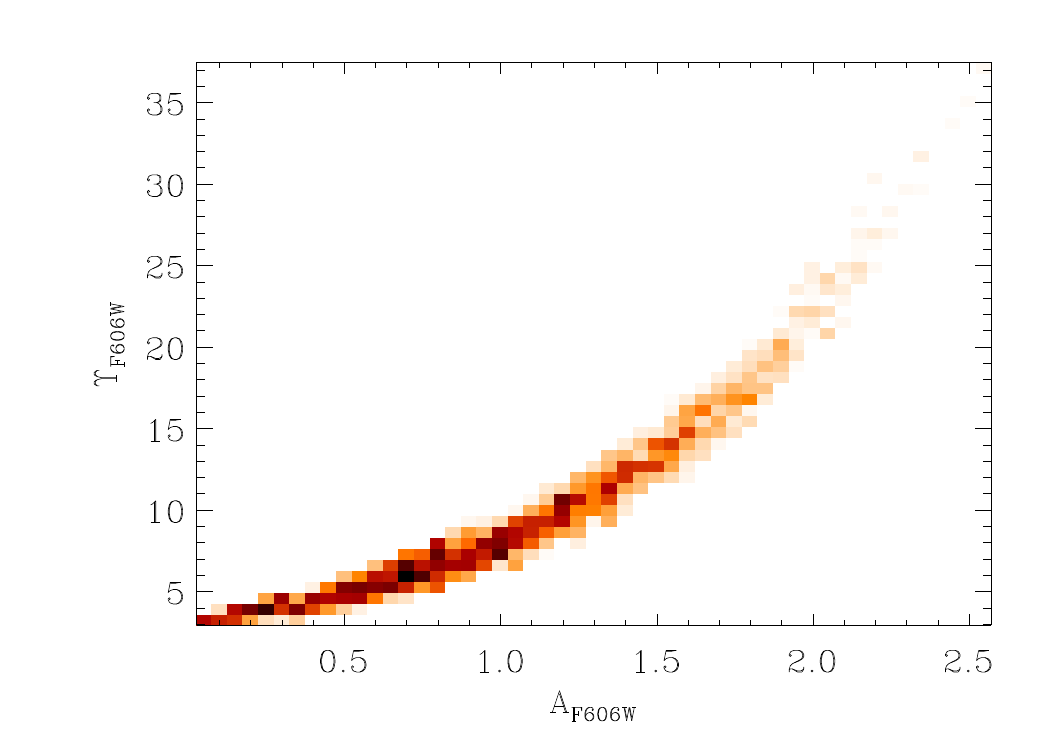}
\caption{
2D histogram showing the joint probability distribution for the stellar mass-to-light ratio, $\ML$, and the amount of extinction in the F606W band, from FSPS models.  As the amount of extinction increases, the model magnitude in the F606W filter increases, leading to a larger values of $\ML$.  Results are shown for models with $h=0.7$ and Solar metallicity, but the distributions for other values of $h$ and metallicity exhibit a similar behavior.}
\label{fig:fsps_dust}
\end{figure}

Figure \ref{fig:h_ml} shows what happens to the posterior probability distribution for the Hubble constant when we impose the SPS constraints on $\ML$.  Using the FSPS models has little effect on the $h$ distribution, because these models allow for a large range of values for $\ML$.  The Maraston models, by contrast, favor $\ML\sim4.5$--6.5 and such values tend to reduce the median $h$ and tighten the distribution.  Clearly it is important to understand why the FSPS and Maraston models differ as to whether high values of $\ML$ are acceptable.  Figure \ref{fig:fsps_dust} shows that the high $\ML$ values attained in FSPS models correspond to large values of extinction --- values that seem surprising for a luminous early-type galaxy in a modest cluster at redshift $z=0.361$.  We infer that the flexibility of FSPS models is allowing them to reproduce the observed colors of the galaxy even with models that do not make much sense astrophysically.  One way to reconcile the FSPS and Maraston models is to constrain the amount of dust in our FSPS models.  We find that adopting an extinction prior of $A_{F606W}=0.45\pm0.2$ would bring the FSPS constraints on $\ML$ into agreement with those from the Maraston models.  Imposing such a prior has little effect on the $\ML$ constraints from the Maraston models since those models show little or no correlation between dust extinction and $\ML$.

Ultimately we need to decide what to use for our final constraints on $\ML$.  Since the Maraston models are constructed to match the SDSS LRG sample and require no ad hoc assumptions about the dust content of the lens galaxy, we elect to use them when reporting our final determination of $h$ and model parameters (see Table \ref{tab:mlresults}).  With these constraints on $\ML$ we find \hml.

\subsection{Physical Properties of NFW Halos}
\label{sec:NFW}

When fitting models with an NFW dark matter halo, we previously took both $r_s$ and $\kappa_s = \rho_s r_s/\Sigma_{crit}$ to be free parameters.  However, in $N$-body simulations the two NFW parameters are actually related to one another, albeit with some scatter.  We now consider whether our lens model parameters have reasonable values in general, and whether they are consistent with the correlation found in simulated halos.

\begin{figure}[b]
\centering
\includegraphics[clip=true, trim=1.35cm 0cm 0cm 0cm,width=9cm]{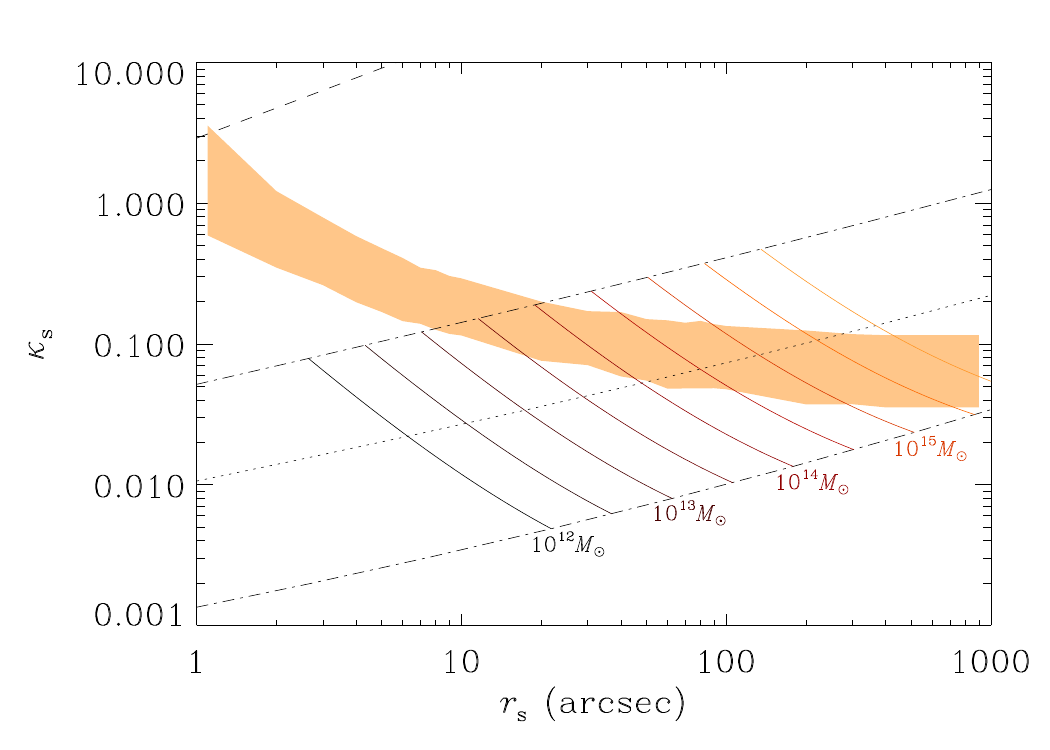}
\caption{The space of scale radius, $r_s$, and halo normalization, $\kappa_s$, for NFW halos at the lens redshift $z=0.361$.  The shaded region indicates the parameter range occupied by our lens models with an NFW dark matter halo.  The dotted curve shows the expected relation between $\kappa_s$ and $r_s$ based on the median concentration/mass relation found in $N$-body simulations by \citet{Maccio07}, while the dot-dash curves show the $\pm3\sigma$ range, and the dashed curve in the upper left corner lies $6\sigma$ above the median.  The colored solid curves represent the theoretical predictions at fixed virial mass, with the concentration varying $\pm3\sigma$ around the median, ranging from $10^{12}\,M_\Sun$ (black) to $10^{16}\,M_\Sun$ (orange) in steps of 0.5 dex.}
\label{fig:nfwthy}
\end{figure}

Parameterizing NFW halos with the virial mass $M_{vir}$ and concentration $c_v$, \citet{Maccio07} find that the parameters are related by $c_v(z) = 213^{+40}_{-34} \,M_{vir}^{-0.109\pm0.005}\,(1+z)^{-1}$.  We can express our lens model parameters in terms of $M_{vir}$ and $c_v$ as follows:
\begin{eqnarray}
\kappa_s &=& \left(\frac{1+z}{c}\right)^2\left[\frac{1}{2}\left(GM_{vir}\right)^{1/2}H_0^2\Omega_m\Delta_{vir}\right]^{2/3} \\ && \times \frac{D_{ol}D_{ls}}{D_{os}}c_v^2\left[\ln(1+c_v)-\frac{c_v}{1+c_v}\right]^{-1} \nonumber \\
r_s &=& \frac{1}{c_v(1+z)}\left[\frac{2GM_{vir}}{H_0^2\Omega_m\Delta_{vir}}\right]^{1/3}
\end{eqnarray}
where $\Delta_{vir}=98$ is the virial overdensity \citep{Mainini, Maccio07}.  Figure \ref{fig:nfwthy} compares our recovered model values of $r_s$ and $\kappa_s$ with expectations based on the \citet{Maccio07} relation for different values of the NFW halo mass.  The first point is that the model parameters do indeed have reasonable values.  Going into more detail, we see that our models with small scale radii ($r_s<10''$) are consistent with relatively low cluster masses ($M_{vir} \lesssim 10^{13.5}\,M_\Sun$) but concentrations that are 3--$6\sigma$ above the median for that mass.  As the scale radius increases, our models follow a track corresponding to increasing mass and decreasing concentration; indeed, lens models with $r_s \gtrsim 200''$ require an extraodinarily large mass of $M_{vir} \gtrsim 10^{15}\,M_\Sun$.  Such a large mass seems unreasonable for a fairly modest galaxy cluster, especially considering that X-ray measurements imply a mass within 1 $h_{75}^{-1}$ Mpc of $9.9^{+1.9}_{-3.8} \times 10^{13}\,M_\Sun$ \citep{Chartas}.  Therefore, we argue that our models with $r_s \gtrsim 200''$ are disfavored but models with $r_s \lesssim 200''$ have parameters that seem reasonable in comparison with simulated NFW halos.  We note that recovering a concentration that is a few sigma above the median is not necessarily a concern, because there may be a selection bias such that a high concentration increases the lensing cross section \citep[see, e.g.,][]{mandelbaum09}.

\section{Discussion and Conclusions}
\label{sec:conclusions}

Since its discovery in 1979, Q0957 has presented many puzzles that we still cannot definitively solve.  Nevertheless, by combining new HST/ACS data and stellar population synthesis models, we have presented a consistent picture of lensing in this system using a realistic treatment of both the stellar and dark matter components of the mass distribution.  In \S \ref{sec:H0} we discuss our results regarding measurements of $H_0$ with Q0957.  Turning the tables, in \S \ref{sec:massdist} we adopt priors on $H_0$ from other measurements and examine the inferred properties of the lens mass distribution.  Looking ahead, in \S \ref{sec:future} we discuss potential ways in which these measurements can be improved.

\subsection{Hubble Constant}
\label{sec:H0}

Motivated by our new ACS data, we conducted a joint strong+weak lensing analysis in the hope of obtaining the best constraints to date from Q0957.  In \S \ref{sec:basic} we found \hacs\ on the basis of lensing alone.  This result is higher than, but still consistent with, measurements from other recent lensing \citep[e.g.,][]{Jakobsson, Paraficz, Oguri} and non-lensing \citep[e.g.,][]{Freedman, Riess, Dunkley, Komatsu} studies.  In spite of the extensive lensing data we have obtained, the uncertainty in $H_0$ from Q0957 is still larger than from most other lenses (see Fig.\ \ref{fig:hlenses}).

One source of uncertainty in Q0957 is the sheer complexity of the potential: with ellipticity, shear, and higher-order environmental terms to play with, models can find a wide range of combinations that fit the data well (see Figs.\ \ref{fig:a102d}--\ref{fig:nfw2d}).  The main systematic effect in lens models is a correlation between $h$ and the stellar mass-to-light ratio of the lens galaxy.  Varying $\ML$ changes the balance between stars and dark matter in the lens galaxy, which modifies the net density profile, which then affects $h$ through the radial profile degeneracy \citep[e.g.,][]{Kochanek02}.  We can actually turn this degeneracy to our advantage if we can place independent constraints on the stellar mass-to-light ratio.  In \S \ref{sec:SPS} we used the stellar population synthesis models from \citet{Maraston08} to constrain $\ML$ and thereby reduce the uncertainties in our Hubble constant determination to \hml.

While this is a significant reduction in the uncertainty for $H_0$ from Q0957, we caution that SPS models are still improving and may ultimately be even more complicated than the Maraston models.  When we used the FSPS models of \citet{Conroy}, for example, we did not see much tightening of the $H_0$ constraints relative to lensing alone; and we traced the trouble to uncertainties in the amount of dust extinction in the lens galaxy.  There is one additional source of uncertainty in SPS models that we did not explicitly address, namely the IMF.  Variations in the IMF can alter the colors and mass-to-light ratios of SPS models \citep[e.g.,][]{Conroy}.  In particular, a more bottom-heavy IMF (e.g., Salpeter) would raise the inferred value of the $\ML$ and hence our median value of $H_0$, while a more top-heavy IMF would have the opposite effect.  For the Maraston SPS models, variations in the IMF must be relatively small or the models would cease to provide a good fit to the SDSS LRG sample.  We attempt to compensate for IMF-related variations by allowing broad ranges for the star formation history, star formation epoch, and dust.  Nevertheless, this remains an unknown, but presumably small, uncertainty in our models.  Clearly there is a lot of room for improvement with a better understanding of the stellar population of the lens galaxy (see \S \ref{sec:future}).

\subsection{Mass Distribution}
\label{sec:massdist}

Instead of trying to measure $H_0$ ourselves, we can choose to place external priors on $H_0$ to see how well we can understand the mass distribution of the lens.  We consider two determinations of $H_0$: the refurbished distance ladder measurement of $H_0=74.2\pm3.6$ \Hunits\ by \citet{Riess}, and the combined WMAP5+SNe+BAO value of $H_0=70.5\pm1.3$ from \citet{Komatsu}.  In Table \ref{tab:acshresults} we show the model parameters recovered from this approach.

Examining the relative probabilities of the models, it is clear that the lensing data favor a power law profile of index $\alpha=0.5$ or 1.0 over $\alpha=1.5$ or an NFW profile.  Both of the favored models exhibit relatively large core radii: $a=5.8^{+1.1}_{-1.0}$ arcsec for $\alpha=0.5$, or $a=4.2^{+1.0}_{-0.7}$ arcsec for $\alpha=1$.  Given the reduced probability of our NFW models, we conclude that lensing provides strong evidence for a constant-density core (rather than a cusp) in the dark matter halo of Q0957.

\begin{figure}[t]
\centering
\includegraphics[clip=true, trim=1.4cm 0cm 0cm 0cm,width=9cm]{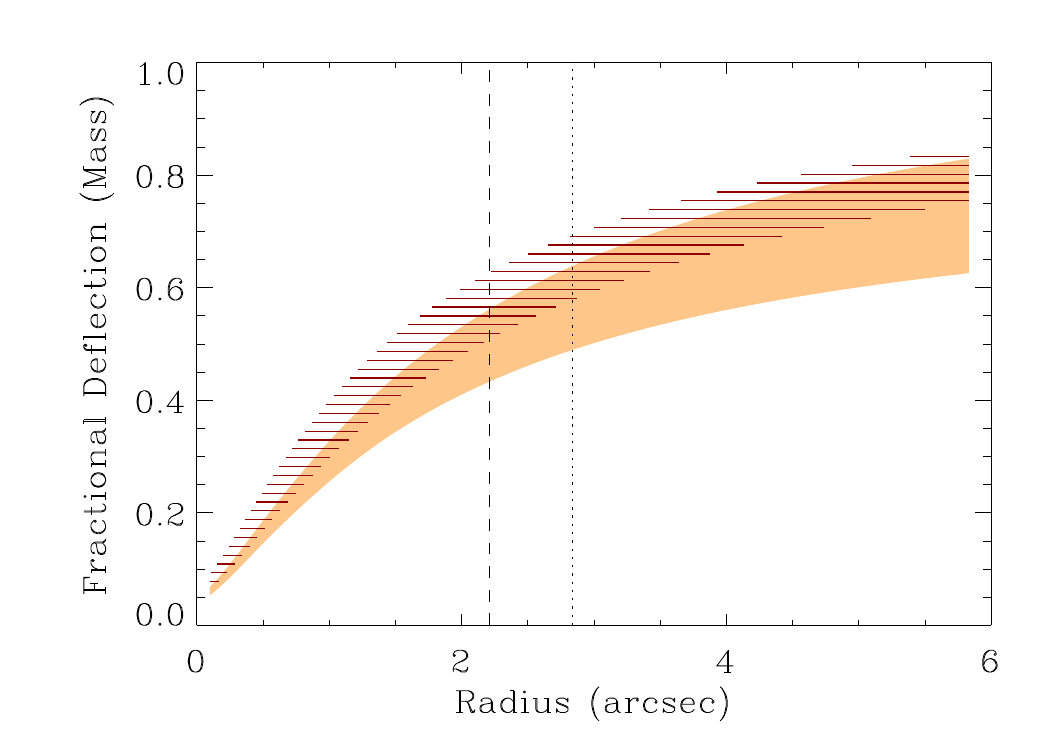}
\caption{Fraction of the monopole deflection contributed by the dark matter halo, as a function of radius.  Since $\alpha(r) \propto M(r)/r$, this plots also depicts the enclosed projected dark matter fraction as a function of radius.  The vertical dotted line indicates the Einstein radius at $\Rein=2.84''$ while the vertical dashed line marks the effective radius of the stellar light profile at $R_e=2.21''$.  The shaded regions show results from our models when we adopt different priors on the Hubble constant: the orange region corresponds to assuming $H_0=74.2\pm3.6\ \Hunits$ from \citet{Riess}, while the region indicated with red horizontal lines corresponds to $H_0=70.5\pm1.3\ \Hunits$ from \citet{Komatsu}.}
\label{fig:frac}
\end{figure}

In Figure \ref{fig:defplot} we showed that Q0957 exhibits a rising deflection profile, analogous to a rising rotation curve and indicative of a net density profile shallower than isothermal.  While this is not the first case of a lens with a rising rotation curve \citep[see][]{Kochanek06}, the origin of phenomenon is unclear.  One possible explanation involves the special location of the lens galaxy.  As the central galaxy in a modest cluster, the lens is embedded in the most concentrated part of a massive dark matter halo.  The higher than average surface density of dark matter leads to a shallow density profile and a rising deflection curve.  To further explore this idea, we shown in Figure \ref{fig:frac} the fraction of the deflection contributed by dark matter as a function of radius.  This is equivalent to the projected enclosed dark matter fraction as a function of radius.  At the effective radius, dark matter constitutes $(50\pm7)\%$ or $(57\pm7)\%$ of the enclosed mass (assuming the Riess or Komatsu priors on $H_0$, respectively).  Such values are well above the minimum dark matter fraction found $(38\pm7)\%$ for galaxies in the SLACS survey \citep{Bolton}, and are greater than 16 of the 22 lens systems considered by \citep{Jiang}.  It is not very surprising, of course, to find a relatively large dark matter fraction in a brightest cluster galaxy.

We find the the dark matter halo in Q0957 must be well aligned with stellar mass distribution.  With either the Riess or Komatsu $H_0$ priors, we find the position angle of the dark matter halo to be $\theta_e=73^{+9}_{-10}$ deg, in good agreement with the measured position angle of the stellar component $\theta_{\star}=71\pm 5$ deg at large radii ($>5''$).   Perhaps more interesting is that the ellipticity of the dark matter appears to be in good agreement with that of the stellar distribution.  We find the ellipticity of the dark matter to be $e=0.28^{+0.06}_{-0.07}$ or $e=0.25^{+0.07}_{-0.09}$ for the Riess or Komatsu $H_0$ values, in good agreement with the measured value of the stellar ellipticity $e_{\star} \sim 0.3$ at large radii.

In general, we find that improving $H_0$ constraints from 5\% to 2\% has little impact on our understanding of the mass distribution, because most of our model parameters have little or no correlation with $H_0$.  The only significant exception is in our determination of the stellar mass to light ratio, $\ML$.  We find $\ML=5.5^{+0.9}_{-0.5}$ using the Riess value for $H_0$, versus $\ML=5.5^{+0.2}_{-0.3}$ using the Komatsu value.  For comparison, the Maraston SPS models give $\ML=5.9\pm1.9$.  It is interesting to see that combining lensing with $H_0$ priors can provide excellent constraints on stellar populations, which may prove useful as SPS laboratories as multi-wavelength datasets for well-studied lenses grow.

We found the intriguing result that our models favor a quasar flux ratio around $F_B/F_A=0.53\pm0.06$, which is substantially different from the radio measurement of $F_B/F_A=0.74\pm0.02$.  Since the radio emission is free from extinction by dust, and should be insensitive to microlensing by stars, we infer the quasar flux ratio in Q0957 seems to be ``anomalous."  The putative anomaly presumably indicates additional complexity in the lens potential.  While the nature of that complexity is not yet clear, it is worth noting that dark matter substructure can easily produce flux ratio anomalies \citep[e.g.,][]{Mao, Metcalf, Dalal, Chiba, Macleod}.  Invoking substructure as an explanation seems especially enticing because the radio jets, a mere 80 mas away from the quasar images, have a flux ratio of $0.61\pm0.04$ \citep{Bonometti}, in much better agreement with the macro models.  While we cannot definitively identify substructure from the present analysis, the evidence is fascinating and warrants future study.

\subsection{Future Prospects}
\label{sec:future}

Looking ahead, there are several ways in which we can hope to improve the measurement of $H_0$ in Q0957.  Following this work, the best improvements are likely to come from a better understanding of the stellar population and $\ML$.  In the near term, an extension of the photometric data to both bluer and redder wavelengths could reduce the uncertainties in existing SPS models and, therefore, the uncertainties in $H_0$.  Over the longer term, extensive photometry spanning UV/optical/infrared wavelengths, coupled with a better understanding of SPS uncertainties (e.g., TP-AGB stars, blue stragglers, IMF), should boost the reliability and reduce the uncertainties in the SPS technique.

Improved lensing constraints may also help.  One source of systematic uncertainty in our models is the uncertainty in the measurement of the total convergence from weak lensing.  Currently, the total convergence is measured to $\sim 34\%$ precision.  Deeper imaging would be observationally expensive but worthwhile, especially if coupled with improved understanding of the point spread function and source redshift distribution for the weak lensing analysis.  For strong lensing, we have noted that there are additional faint lensed features that we have not used, but that might further constrain the lensing critical curve (see the discussion accompanying Figs.\ \ref{fig:crit} and \ref{fig:fourimage}).  We note for the record that using the precise position and flux ratio constraints for the quasars is not likely to help us understand the (large-scale) mass distribution or $H_0$ any better; those data will ultimately be most useful for probing small-scale structure in the lens.

Finally, it is interesting to consider whether stellar dynamics data could further constrain our models.  Previous studies have shown that joint lensing+dynamics analyses can provide more information than lensing alone about the mass distribution \citep[e.g.,][]{Treu, Barnabe07}.  The joint approach has been used to good effect by the SLACS team to improve constraints on quantities like the total mass to light ratio and the slope of the inner density profile \citep{Koopmans}.  For Q0957, \citet{Romanowsky} have successfully combined information from stellar dynamics and lensing to tighten constraints on mass models.  Adopting the measured central velocity dispersion from \citet{Tonry}, \citeauthor{Romanowsky} use orbit modeling techniques to constrain a spherical power law profile, measuring $H_0=61^{+13}_{-15}$ \Hunits\ ($2\sigma$). While it is clear such an analysis would be useful to further constrain our models, it would require orbit modeling for each of our model classes, which is beyond the scope of this work.  Furthermore, it would require a deprojection of the stellar component that we include in our models, which could be challenging (because of the varying ellipticity and orientation; cf.\ Figure \ref{fig:e-and-pa}) and would have uncertainties of its own.


\acknowledgements
We thank Claudia Maraston and Charlie Conroy for providing their SPS models as well as technical comments and assistance, Chien Peng for revised NICMOS photometry and helpful discussions, Andrew Baker and Chuck Joseph for helpful comments and suggestions, and Viviana Acquaviva for timely help and advice.  We also thank the anonymous referee for constructive criticism.  This work has been supported by grant HST-GO-10569 from the Space Telescope Science Institute, which is operated by the Association of Universities for Research in Astronomy, Inc., under NASA contract NAS5-26555.  Additional support has come from the National Science Foundation via grants AST-0747311 (RF and CRK) and AST-0607667 (RN and GMB).



\clearpage

\begin{deluxetable}{cccc}
\tablewidth{0pt}
\tablecaption{Lens Galaxy Photometry}
\tablehead{
Filter & Total Counts & $m_{AB}$ & Zeropoint, AB\tablenotemark{1}}
\startdata
F606W & 3328.6 & $18.809 \pm 0.061$ & 27.614 \\
F814W & 5372.9 & $17.743 \pm 0.065$ & 27.068 \\
\enddata
\tablenotetext{1}{Quoted zeropoints differ from standard values as we must correct for a plate scale change from $0.05''$ to $0.03''$ pixels}
\label{tab:galphot}
\end{deluxetable}

\begin{deluxetable}{ccc}
\tablewidth{0pt}
\tablecaption{Modeling Constraints}
\tablehead{
Feature & Position $(\arcsec)\tablenotemark{1} $ & Symbol in Figure 3
}
\startdata
G1 & $(0,0)\pm0.00001$ & $-$ \\
Quasar A & $(1.408,5.034)\pm0.03$ & Green Plus \\
Quasar B & $(0.182,-1.018)\pm0.03$ & Green Plus \\
Jet A5 & $(1.392,5.080)\pm0.03$ & $-$ \\
Jet B5 & $(0.164,-0.962)\pm0.03$ & $-$ \\
IA & $(2.878,3.453)\pm0.05$ & Red Circle \\
IB & $(-1.362,-0.043)\pm0.05$ & Red Circle \\
IIA & $(2.666,3.634)\pm0.05$ & Blue Circle \\
IIB & $(-1.457,-0.075)\pm0.05$ & Blue Circle \\
IIIA & $(2.395,3.694)\pm0.05$ & Cyan Circle \\
IIIB & $(-1.682,-0.026)\pm0.05$ & Cyan Circle \\
IVA & $(0.021,-2.532)\pm0.03$ & Green Diamond \\
IVB & $(0.512,-2.386)\pm0.03$ & Green Diamond \\
VA & $(2.111,3.664)\pm0.05$ & Red Diamond \\
VB & $(-0.768,-2.640)\pm0.05$ & Red Diamond \\
VIA & $(1.875,3.488)\pm0.12$ & Blue Diamond \\
VIB & $(-1.128,-2.777)\pm0.05$ & Blue Diamond \\
VIC & $(1.523,-1.634)\pm0.05$ & Blue Diamond \\
VIIA & $(1.875,3.488)\pm0.12$ & $-$ \\
VIIB & $(-1.065,-2.786)\pm0.05$ & $-$ \\
VIIC & $(1.489,-1.688)\pm0.05$ & $-$ \\
VIIIA & $(2.280,3.391)\pm0.09$ & Magenta Diamond \\
VIIIB & $(-0.454,-2.873)\pm0.09$ & Magenta Diamond \\
IXA & $(-2.003,-2.435)\pm0.05$ & Cyan Diamond \\
IXB & $(1.293,3.611)\pm0.05$ & Cyan Diamond \\
XA & $(-1.708,-1.878)\pm0.08$ & Yellow Circle \\
XB & $(-2.181,-0.551)\pm0.08$ & Yellow Circle \\
XIA & $(-2.070,-1.252)\pm0.08$ & Magenta Circle \\
XIB & $(-2.112,-1.053)\pm0.08$ & Magenta Circle \\
XIIA & $(-2.745,-0.516)\pm0.08$ & Green Circle \\
XIIB & $(-2.781,-0.699)\pm0.08$ & Green Circle \\
\enddata
\tablecomments{New features are indicated with Roman numerals}
\tablenotetext{1}{Written as $(x,y)$ where $x$ is West and $y$ is North.}
\label{tab:data}
\end{deluxetable}

\begin{deluxetable}{lc}
\tablewidth{0pt}
\tablecaption{Model Parameters}
\tablehead{
  Parameter & Label 
}
\startdata
Stellar mass to light ratio & $\ML$ \\
Halo ellipticity & $e$  \\
Position angle & $PA$  \\
Shear angle & $\theta_{\gamma}$  \\
Core radius ($\alpha$ models) & $a$  \\
Scale radius (NFW models) & $r_s$  \\
$\sigma$ angle & $\theta_{\sigma}$  \\
$\delta$ angle & $\theta_{\delta}$  \\
Halo mass normalization & $b_d$ \\
External Shear & $\gamma$  \\
$3^{\rm{rd}}$ order term & $\sigma$  \\
$3^{\rm{rd}}$ order term & $\delta$ \\
\enddata
\tablecomments{Power law $\alpha$ models use a core radius $a$, while NFW models use a scale radius $r_s$.}
\label{tab:parms}
\end{deluxetable}

\begin{deluxetable}{lccc}
\tablewidth{0pt}
\tablecaption{SPS Model Parameters}
\tablehead{
  Parameter & Prior & FSPS & Maraston SPS
}
\startdata
Formation redshift, $z_f$ & $0.361 - 1089$ & \checkmark & \checkmark \\
SFR e-folding time, $\tau$ & $0 - \infty$ & \checkmark & \checkmark \\
Constant SFR, $C$ & $0 - 1$ & \checkmark & \checkmark \\
Dust around young stars, $\tau_1$ & $0 - \infty$ & \checkmark & \checkmark \\
Diffuse dust\tablenotemark{1}, $\tau_2$ & $P \propto e^{-1.086\tau_2}$ & \checkmark & \checkmark \\
Fraction of blue HB stars\tablenotemark{2}, $f_{BHB}$ & $0 - 0.5$ & \checkmark & $-$ \\
Specific frequency of blue stragglers\tablenotemark{2}, $S_{\rm{BS}}$ & $0 - 10$ & \checkmark & $-$ \\
Shift in $log(L_{bol})$ along the TP-AGB\tablenotemark{2}, $\Delta_{L}$ & $-0.4$ $-$ 0.4  & \checkmark & $-$ \\
Shift in $log(T_{eff})$ along the TP-AGB\tablenotemark{2}, $\Delta_{T}$ & $-0.2$ $-$ 0.2  & \checkmark & $-$ \\
\enddata
\tablenotetext{1}{We use a exponential prior on the diffuse dust content of the form $e^{-A_{F606W}/1.0 \rm{mag}}$}
\tablenotetext{2}{See \citet{Conroy} for details}
\label{tab:spsparms}
\end{deluxetable}

\begin{deluxetable}{lccccccccc}
\tablewidth{0pt}
\tablecaption{Model Results: HST-ACS data}
\tablehead{
\multicolumn{10}{c}{Median model values with 68\% CL (95\% CL) uncertainties} \\
\multicolumn{2}{c}{Parameter} & \multicolumn{2}{c}{NFW} & \multicolumn{2}{c}{$\alpha=0.5$} & \multicolumn{2}{c}{$\alpha=1.0$} & \multicolumn{2}{c}{$\alpha=1.5$}
}
\startdata
\multicolumn{2}{c}{$\ML$} & $7.7_{-2.7}^{+3.3}$ & $(_{-4.1}^{+4.9})$ &  $5.4_{-0.6}^{+2.5}$ & $(_{-1.2}^{+3.9})$ & $7.4_{-1.8}^{+1.5}$ & $(_{-2.7}^{+3.0})$ & $8.6_{-1.5}^{+1.1}$ & $(_{-2.6}^{+2.2})$ \\
\multicolumn{2}{c}{$b_d$ ($''$)} & $0.097_{-0.022}^{+0.01}$ & $(_{-0.034}^{+0.118})$  & $6.3_{-1.4}^{+1.6}$ & $(_{-2.3}^{+2.6})$ & $4.4_{-0.8}^{+1.5}$ & $(_{-1.5}^{+2.6})$ & $3.1_{-0.5}^{+0.6}$ & $(_{-0.8}^{+1.0})$ \\
\multicolumn{2}{c}{$e$} & $0.46_{-0.19}^{+0.19}$ & $(_{-0.28}^{+0.30})$ & $0.31_{-0.13}^{+0.14}$ & $(_{-0.18}^{+0.22})$ & $0.40_{-0.14}^{+0.12}$ & $(_{-0.22}^{+0.24})$ & $0.44_{-0.11}^{+0.15}$ & $(_{-0.17}^{+0.24})$ \\
\multicolumn{2}{c}{$\theta_e$ ($^\circ$)} & $70_{-6}^{+7}$ & $(_{-11}^{+12})$ & $72_{-10}^{+9}$ & $(_{-154}^{+14})$ & $71_{-7}^{+9}$ & $(_{-18}^{+14})$ & $72_{-6}^{+9}$ & $(_{-11}^{+14})$ \\
\multicolumn{2}{c}{$\gamma$ ($\times10^{2}$)} & $7.8_{-2.7}^{+3.0}$ & $(_{-4.3}^{+5.0})$ & $7.6_{-2.8}^{+2.1}$ & $(_{-4.0}^{+3.5})$ & $7.8_{-2.5}^{+3.1}$ & $(_{-4.0}^{+5.2})$ & $7.5_{-3.4}^{+4.2}$ & $(_{-5.2}^{+5.8})$ \\
\multicolumn{2}{c}{$\theta_{\gamma}$ ($^\circ$)} & $52_{-15}^{+11}$ & $(_{-25}^{+15})$ & $51_{-13}^{+11}$ & $(_{-20}^{+16})$ & $51_{-14}^{+10}$ & $(_{-25}^{+15})$ & $42_{-19}^{+16}$ & $(_{-33}^{+26})$ \\
\multicolumn{2}{c}{$a$ ($''$)} & $-$ & $-$ & $5.7_{-1.1}^{+1.2}$ & $(_{-2.0}^{+2.0})$ & $4.0_{-0.9}^{+1.0}$ & $(_{-1.4}^{+1.7})$ & $1.8_{-0.7}^{+0.5}$ & $(_{-1.0}^{+1.0})$ \\
\multicolumn{2}{c}{$\sigma$ ($\times10^{3}$)} & $2.6_{-1.3}^{+1.7}$ & $(_{-1.9}^{+2.7})$ & $1.9_{-0.9}^{+1.1}$ & $(_{-1.3}^{+1.8})$ & $1.8_{-0.9}^{+1.3}$ & $(_{-1.3}^{+2.0})$ & $1.9_{-1.4}^{+1.0}$ & $(_{-1.5}^{+1.6})$ \\
\multicolumn{2}{c}{$\theta_{\sigma}$ ($^\circ$)} & $335_{-291}^{+5}$ & $(_{-322}^{+14})$ & $210_{-129}^{+46}$ & $(_{-180}^{+95})$ & $233_{-184}^{+61}$ & $(_{-218}^{+106})$ & $266_{-197}^{+51}$ & $(_{-242}^{+85})$ \\
\multicolumn{2}{c}{$\delta$ ($\times10^{3}$)} & $2.8_{-1.1}^{+1.8}$ & $(_{-1.8}^{+3.0})$ & $2.6_{-1.0}^{+1.0}$ & $(_{-1.6}^{+1.8})$ & $2.4_{-1.0}^{+1.5}$ & $(_{-1.7}^{+2.6})$ & $3.3_{-1.4}^{+0.9}$ & $(_{-2.0}^{+1.9})$ \\
\multicolumn{2}{c}{$\theta_{\delta}$ ($^\circ$)} & $68_{-17}^{+9}$ & $(_{-28}^{+17})$ & $61_{-13}^{+11}$ & $(_{-21}^{+18})$ & $59_{-13}^{+17}$ & $(_{-21}^{+21})$ & $62_{-5}^{+10}$ & $(_{-18}^{+15})$ \\
\hline
\multicolumn{2}{c}{$f_B/f_A$} & $0.55_{-0.06}^{+0.10}$ & $(_{-0.08}^{+0.22})$ & $0.53_{-0.05}^{+0.05}$ & $(_{-0.08}^{+0.09})$ & $0.50_{-0.04}^{+0.06}$ & $(_{-0.07}^{+0.09})$ & $0.51_{-0.04}^{+0.07}$ & $(_{-0.06}^{+0.12})$ \\

\multicolumn{2}{c}{$\kappa_{s,30''}$} & $0.12_{-0.02}^{+0.07}$ & $(_{-0.05}^{+0.10})$ & $0.15_{-0.05}^{+0.07}$ & $(_{-0.08}^{+0.13})$ & $0.21_{-0.07}^{+0.07}$ & $(_{-0.10}^{+0.11})$ & $0.27_{-0.04}^{+0.04}$ & $(_{-0.06}^{+0.07})$ \\
\multicolumn{2}{c}{$H_0$} & $89.4_{-13.7}^{+16.8}$ & $(_{-20.8}^{+27.3})$ & $75.3_{-10.0}^{+13.8}$ & $(_{-15.3}^{+23.4})$ &  $84.4_{-11.3}^{+11.1}$ & $(_{-17.9}^{+18.3})$ & $89.9_{-10.0}^{+10.1}$ & $(_{-16.0}^{+17.0})$ \\
\hline
\multicolumn{2}{c}{min$(\chi^2_{reduced}$)} & \multicolumn{2}{c}{0.38} & \multicolumn{2}{c}{0.91} & \multicolumn{2}{c}{0.90} & \multicolumn{2}{c}{0.58}\\
\multicolumn{2}{c}{$P_{rel}$} & \multicolumn{2}{c}{0.014} & \multicolumn{2}{c}{0.575} & \multicolumn{2}{c}{1.000} & \multicolumn{2}{c}{0.114}\\
\enddata
\label{tab:acsresults}
\end{deluxetable}

\begin{deluxetable}{lccccccccc}
\tablewidth{0pt}
\tablecaption{Model Results: HST-ACS data + Maraston SPS Models}
\tablehead{
\multicolumn{10}{c}{Median model values with 68\% CL (95\% CL) uncertainties} \\
\multicolumn{2}{c}{Parameter} & \multicolumn{2}{c}{NFW} & \multicolumn{2}{c}{$\alpha=0.5$} & \multicolumn{2}{c}{$\alpha=1.0$} & \multicolumn{2}{c}{$\alpha=1.5$}
}
\startdata
\multicolumn{2}{c}{$\ML$} & $5.7_{-1.2}^{+1.2}$ & $(_{-1.9}^{+1.9})$ &  $5.3_{-0.9}^{+1.1}$ & $(_{-1.5}^{+2.9})$ & $6.5_{-1.8}^{+1.0}$ & $(_{-2.0}^{+1.6})$ & $6.9_{-1.0}^{+0.9}$ & $(_{-1.4}^{+1.4})$ \\
\multicolumn{2}{c}{$b_d$ ($''$)} & $0.118_{-0.018}^{+0.028}$ & $(_{-0.025}^{+0.107})$  & $6.4_{-1.2}^{+1.5}$ & $(_{-2.0}^{+2.5})$ & $5.0_{-2.0}^{+1.3}$ & $(_{-1.8}^{+2.3})$ & $3.5_{-0.5}^{+0.5}$ & $(_{-0.9}^{+0.9})$ \\
\multicolumn{2}{c}{$e$} & $0.34_{-0.09}^{+0.09}$ & $(_{-0.26}^{+0.14})$ & $0.30_{-0.11}^{+0.6}$ & $(_{-0.16}^{+0.16})$ & $0.32_{-0.14}^{+0.10}$ & $(_{-0.16}^{+0.115})$ & $0.34_{-0.08}^{+0.08}$ & $(_{-0.12}^{+0.11})$ \\
\multicolumn{2}{c}{$\theta_e$ ($^\circ$)} & $70_{-8}^{+8}$ & $(_{-13}^{+15})$ & $73_{-10}^{+8}$ & $(_{-154}^{+13})$ & $72_{-9}^{+9}$ & $(_{-15}^{+13})$ & $76_{-6}^{+8}$ & $(_{-11}^{+12})$ \\
\multicolumn{2}{c}{$\gamma$ ($\times10^{2}$)} & $8.6_{-2.9}^{+3.0}$ & $(_{-4.4}^{+4.5})$ & $7.7_{-2.1}^{+2.8}$ & $(_{-3.4}^{+4.1})$ & $8.2_{-2.5}^{+2.8}$ & $(_{-4.2}^{+5.1})$ & $9.8_{-3.4}^{+2.9}$ & $(_{-5.1}^{+6.1})$ \\
\multicolumn{2}{c}{$\theta_{\gamma}$ ($^\circ$)} & $55_{-12}^{+8}$ & $(_{-19}^{+13})$ & $51_{-14}^{+10}$ & $(_{-20}^{+16})$ & $52_{-12}^{+9}$ & $(_{-21}^{+15})$ & $49_{-9}^{+7}$ & $(_{-17}^{+15})$ \\
\multicolumn{2}{c}{$a$ ($''$)} & $-$ & $-$ & $5.8_{-1.1}^{+1.1}$ & $(_{-1.9}^{+1.9})$ & $4.0_{-0.9}^{+1.0}$ & $(_{-1.5}^{+1.7})$ & $1.4_{-0.5}^{+0.5}$ & $(_{-0.8}^{+0.8})$ \\
\multicolumn{2}{c}{$\sigma$ ($\times10^{3}$)} & $2.0_{-1.1}^{+1.5}$ & $(_{-1.5}^{+2.5})$ & $1.9_{-1.0}^{+1.1}$ & $(_{-1.3}^{+1.7})$ & $1.8_{-0.9}^{+1.4}$ & $(_{-1.3}^{+2.0})$ & $1.9_{-1.2}^{+1.0}$ & $(_{-1.5}^{+1.6})$ \\
\multicolumn{2}{c}{$\theta_{\sigma}$ ($^\circ$)} & $235_{-95}^{+39}$ & $(_{-174}^{+80})$ & $211_{-119}^{+39}$ & $(_{-162}^{+81})$ & $228_{-96}^{+36}$ & $(_{-195}^{+82})$ & $229_{-70}^{+26}$ & $(_{-146}^{+62})$ \\
\multicolumn{2}{c}{$\delta$ ($\times10^{3}$)} & $2.8_{-1.4}^{+1.4}$ & $(_{-1.9}^{+2.4})$ & $2.5_{-0.9}^{+1.1}$ & $(_{-1.5}^{+1.8})$ & $2.4_{-1.2}^{+1.1}$ & $(_{-1.9}^{+1.9})$ & $3.0_{-1.3}^{+1.2}$ & $(_{-2.0}^{+2.8})$ \\
\multicolumn{2}{c}{$\theta_{\delta}$ ($^\circ$)} & $68_{-16}^{+10}$ & $(_{-28}^{+18})$ & $61_{-12}^{+11}$ & $(_{-21}^{+17})$ & $60_{-14}^{+16}$ & $(_{-23}^{+23})$ & $67_{-14}^{+10}$ & $(_{-23}^{+19})$ \\
\hline
\multicolumn{2}{c}{$f_B/f_A$} & $0.61_{-0.06}^{+0.08}$ & $(_{-0.09}^{+0.18})$ & $0.53_{-0.05}^{+0.05}$ & $(_{-0.07}^{+0.09})$ & $0.53_{-0.06}^{+0.05}$ & $(_{-0.07}^{+0.08})$ & $0.58_{-0.06}^{+0.07}$ & $(_{-0.09}^{+0.11})$ \\

\multicolumn{2}{c}{$\kappa_{s,30''}$} & $0.17_{-0.05}^{+0.03}$ & $(_{-0.09}^{+0.05})$ & $0.13_{-0.02}^{+0.03}$ & $(_{-0.04}^{+0.04})$ & $0.16_{-0.03}^{+0.02}$ & $(_{-0.05}^{+0.04})$ & $0.23_{-0.02}^{+0.01}$ & $(_{-0.02}^{+0.02})$ \\
\multicolumn{2}{c}{$H_0$} & $82.9_{-7.3}^{+6.9}$ & $(_{-12.6}^{+12.2})$ & $73.7_{-7.6}^{+7.6}$ & $(_{-11.9}^{+12.4})$ & $78.4_{-7.6}^{+6.6}$ & $(_{-12.6}^{+10.2})$ & $82.2_{-6.2}^{+5.7}$ & $(_{-10.4}^{+8.7})$ \\
\hline
\multicolumn{2}{c}{min$(\chi^2_{reduced}$)} & \multicolumn{2}{c}{0.57} & \multicolumn{2}{c}{0.99} & \multicolumn{2}{c}{0.98} & \multicolumn{2}{c}{0.83}\\
\multicolumn{2}{c}{$P_{rel}$} & \multicolumn{2}{c}{0.010} & \multicolumn{2}{c}{0.814} & \multicolumn{2}{c}{1.000} & \multicolumn{2}{c}{0.053}\\
\enddata
\label{tab:mlresults}
\end{deluxetable}

\begin{deluxetable}{lccccccccc}
\tablewidth{0pt}
\tablecaption{Model Results: HST-ACS data + $H_0$ priors}
\tablehead{
\multicolumn{10}{c}{Median model values with 68\% CL (95\% CL) uncertainties} \\
\multicolumn{2}{c}{Parameter} & \multicolumn{2}{c}{NFW} & \multicolumn{2}{c}{$\alpha=0.5$} & \multicolumn{2}{c}{$\alpha=1.0$} & \multicolumn{2}{c}{$\alpha=1.5$}
}
\startdata
\multicolumn{2}{c}{$\ML$} & $4.3_{-1.2}^{+0.7}$ & $3.6_{-0.6}^{+0.2}$ &  $5.5_{-0.5}^{+0.5}$ & $5.1_{-0.3}^{+0.2}$ & $5.6_{-0.6}^{+1.3}$ & $5.0_{-0.3}^{+0.2}$ & $5.8_{-0.5}^{+0.5}$ & $5.7_{-0.4}^{+0.3}$ \\
\multicolumn{2}{c}{$b_d$ ($''$)} & $0.135_{-0.018}^{+0.028}$ & $0.144_{-0.011}^{+0.037}$  & $6.3_{-1.0}^{+1.3}$ & $6.7_{-1.1}^{+1.2}$ & $5.5_{-1.1}^{+1.1}$ & $5.6_{-1.5}^{+1.2}$ & $4.1_{-0.5}^{+0.5}$ & $4.1_{-0.5}^{+0.5}$ \\
\multicolumn{2}{c}{$e$} & $0.25_{-0.07}^{+0.06}$ & $0.19_{-0.04}^{+0.05}$ & $0.31_{-0.10}^{+0.07}$ & $0.26_{-0.09}^{+0.07}$ & $0.28_{-0.07}^{+0.08}$ & $0.21_{-0.06}^{+0.08}$ & $0.25_{-0.07}^{+0.06}$ & $25_{-0.06}^{+0.07}$ \\
\multicolumn{2}{c}{$\theta_e$ ($^\circ$)} & $69_{-9}^{+12}$ & $69_{-11}^{+11}$ & $74_{-10}^{+9}$ & $74_{-10}^{+9}$ & $72_{-9}^{+9}$ & $72_{-8}^{+8}$ & $77_{-10}^{+9}$ & $77_{-8}^{+7}$ \\
\multicolumn{2}{c}{$\gamma$ ($\times10^{2}$)} & $9.6_{-2.6}^{+3.0}$ & $10.0_{-2.0}^{+2.0}$ & $7.6_{-1.9}^{+2.7}$ & $7.8_{-2.0}^{+2.4}$ & $8.2_{-2.6}^{+2.5}$ & $8.6_{-2.6}^{+2.2}$ & $11.0_{-3.0}^{+2.3}$ & $11.0_{-2.2}^{+3.0}$ \\
\multicolumn{2}{c}{$\theta_{\gamma}$ ($^\circ$)} & $57_{-11}^{+6}$ & $58_{-8}^{+6}$ & $50_{-14}^{+11}$ & $52_{-14}^{+8}$ & $52_{-12}^{+10}$ & $56_{-10}^{+5}$ & $51_{-8}^{+16}$ & $51_{-7}^{+10}$ \\
\multicolumn{2}{c}{$a$ ($''$)} & $-$ & $-$ & $5.8_{-1.0}^{+1.1}$ & $5.9_{-1.0}^{+1.1}$ & $4.2_{-0.7}^{+1.0}$ & $4.1_{-0.6}^{+0.8}$ & $1.3_{-0.7}^{+0.6}$ & $1.3_{-0.5}^{+0.5}$ \\
\multicolumn{2}{c}{$\sigma$ ($\times10^{3}$)} & $2.5_{-1.1}^{+1.6}$ & $3.3_{-1.4}^{+1.2}$ & $1.9_{-1.0}^{+1.1}$ & $2.1_{-1.1}^{+0.9}$ & $1.9_{-1.1}^{+1.4}$ & $2.0_{-1.0}^{+1.5}$ & $2.2_{-0.9}^{+0.9}$ & $2.2_{-0.9}^{+0.9}$ \\
\multicolumn{2}{c}{$\theta_{\sigma}$ ($^\circ$)} & $213_{-42}^{+23}$ & $211_{-36}^{+10}$ & $215_{-127}^{+45}$ & $215_{-114}^{+30}$ & $221_{-74}^{+24}$ & $220_{-26}^{+21}$ & $180_{-66}^{+47}$ & $185_{-60}^{+39}$ \\
\multicolumn{2}{c}{$\delta$ ($\times10^{3}$)} & $2.8_{-1.5}^{+1.1}$ & $2.4_{-1.2}^{+1.4}$ & $2.3_{-0.7}^{+1.3}$ & $2.5_{-1.6}^{+1.1}$ & $2.3_{-1.5}^{+1.1}$ & $2.3_{-1.3}^{+1.0}$ & $3.8_{-1.3}^{+2.0}$ & $3.8_{-1.0}^{+1.1}$ \\
\multicolumn{2}{c}{$\theta_{\delta}$ ($^\circ$)} & $68_{-16}^{+10}$ & $70_{-28}^{+19}$ & $61_{-12}^{+11}$ & $62_{-16}^{+11}$ & $63_{-16}^{+15}$ & $63_{-16}^{+17}$ & $69_{-10}^{+4}$ & $68_{-8}^{+4}$ \\
\hline
\multicolumn{2}{c}{$f_B/f_A$} & $0.66_{-0.06}^{+0.09}$ & $73_{-0.07}^{+0.22}$ & $0.52_{-0.04}^{+0.05}$ & $0.53_{-0.04}^{+0.05}$ & $0.53_{-0.04}^{+0.05}$ & $0.55_{-0.03}^{+0.05}$ & $0.61_{-0.05}^{+0.04}$ & $0.60_{-0.04}^{+0.05}$ \\

\multicolumn{2}{c}{$\kappa_{s,30''}$} & $0.17_{-0.05}^{+0.04}$ & $0.17_{-0.04}^{+0.03}$ & $0.13_{-0.02}^{+0.02}$ & $0.13_{-0.02}^{+0.02}$ & $0.17_{-0.02}^{+0.02}$ & $0.17_{-0.02}^{+0.02}$ & $0.24_{-0.01}^{+0.01}$ & $0.24_{-0.01}^{+0.01}$ \\
\hline
\multicolumn{2}{c}{$P_{rel}$} & 0.007 & 0.003 & 1.000 & 1.000 & 0.869 & 0.565 & 0.006 & 0.000\\
\enddata
\tablecomments{ Columns on the left use $H_0=74.2\pm3.6$ \Hunits\ from \citet{Riess} while columns on the right use $H_0=70.5\pm1.3$ \Hunits\ from \citet{Komatsu}.}
\label{tab:acshresults}
\end{deluxetable}

\end{document}